\documentclass[journal]{IEEEtran}

\usepackage{amsmath, amssymb, amsthm}
\usepackage{algorithm}
\usepackage{algpseudocode}
\usepackage{graphicx}
\usepackage{subcaption}
\usepackage{setspace}
\usepackage{multicol}
\usepackage[font=small,labelfont=bf]{caption}
\usepackage{listings}
\usepackage{verbatim}
\usepackage{fancyvrb}
\usepackage{xcolor}
\usepackage{adjustbox} 
\usepackage{multirow}
\usepackage{booktabs}
\usepackage[normalem]{ulem}
\usepackage[utf8]{inputenc}
\DeclareUnicodeCharacter{221E}{\ensuremath{\infty}}
\usepackage[hidelinks]{hyperref}
\hypersetup{
    colorlinks=true,
    linkcolor=blue,
    citecolor=blue,
    urlcolor=blue
}

\begin{document}

\title{Multi-LLM Energy Reasoning for Binary-Free Zero-Day IoT Detection}
\author{
Saeid~Jamshidi,
~Foutse~Khomh,
Kawser~Wazed~Nafi,
Omar~Abdul-Wahab,
and Martine~Bella\"{\i}che
\thanks{Corresponding author: Saeid Jamshidi (saeid.jamshidi@polymtl.ca).}%
\thanks{Kawser Wazed Nafi and Foutse Khomh are with the SWAT Laboratory, Polytechnique Montréal, Québec, Canada.}%
\thanks{Saeid Jamshidi, Omar Abdul-Wahab, and Martine Bella\"{\i}che are with the Department of Computer and Software Engineering, Polytechnique Montréal, Québec, Canada.}%
}

\maketitle

\begin{abstract}
Securing Internet of Things (IoT) firmware remains inherently challenging due to proprietary binaries, stripped symbols, heterogeneous hardware architectures, and restricted access to executable code. Established analysis techniques, including static analysis, symbolic execution, and fuzzing, rely on binary transparency together with accurate emulation. When firmware images are encrypted and only partially accessible, the reliability of these methods declines markedly. To address this structural constraint, we introduce a binary-free, architecture-agnostic system that estimates the likelihood of conceptual zero-day exploits from high-level firmware descriptors. The system employs a tri-LLM reasoning architecture comprising an LLaMA 3-8B configuration interpreter, a DeepSeek-v3 structural abstraction analyzer, and a GPT-4o semantic fusion module. Each component provides a distinct analytical perspective, enabling synthesis of configuration semantics, structural patterns, and cross-layer dependencies without reliance on executable artifacts. The system also captures computational signatures produced during LLM inference, including latency dynamics, uncertainty indicators, reasoning-depth traces, and an energy-aware symbolic load formulation. Simulation-based evaluation shows that incremental descriptor perturbations produce a consistent 20--35\% increase in predicted conceptual zero-day likelihood across models. Among the evaluated systems, GPT-4o exhibits stronger cross-layer sensitivity to descriptor variation. Statistical analysis further shows that divergence measures and energy-correlated signals are significantly associated with elevated conceptual risk ($p < 0.01$), supporting descriptor-level triage under binary opacity without claiming to detect exploits.
\end{abstract}

\section{Introduction}
\label{Introduction}
The security of Internet of Things (IoT) firmware has become a critical priority as billions of embedded devices operate autonomously in sensitive, remote, and unmonitored environments \cite{ul2023survey}\cite{el2022secure}\cite{brightwood2024importance}. Firmware governs privileged device functions, including hardware access, authentication, protocol handling, and system initialization, making it central to device integrity \cite{marchand2025firmware}\cite{gujjula2024firmware}\cite{bakhshi2024review}. A single unnoticed flaw can grant adversaries persistent footholds, enable large-scale botnets, disrupt industrial processes, and support stealthy lateral movement across interconnected infrastructures \cite{asadi2024botnets}\cite{almazarqi2024profiling}. As IoT deployments expand in heterogeneity and complexity, firmware security has become essential to cyber resilience at scale \cite{qudus2025advancing}\cite{adewuyi2024convergence}.\\
Despite its importance, firmware analysis remains difficult in practice \cite{gujjula2024firmware}\cite{wu2024your}. IoT vendors frequently employ encrypted binaries, proprietary compression formats, stripped symbols, undocumented architectures, and anti-analysis protections that make disassembly, symbolic execution, fuzzing, and sandboxing unreliable \cite{johnson2025designing}\cite{gomes2025static}. Many devices also lack reliable update mechanisms, leaving long-lived vulnerabilities deployed at scale. Firmware diversity across vendors and hardware families further complicates automated tooling and restricts access to execution semantics \cite{bailey2025symbolic}. Consequently, existing techniques struggle to identify deeply embedded weaknesses, especially zero-day exposure that arises from conceptual design flaws rather than known signatures \cite{karaca2025systematic}.\\
Large Language Models (LLMs) introduce a complementary opportunity for firmware security under limited visibility. In contrast to conventional tools that depend on byte-level access, LLMs support semantic reasoning, hierarchical abstraction, and interpretation of incomplete high-level descriptions \cite{huang2025foundation} \cite{jamshidi2025role} \cite{andreoni2024enhancing}. By analyzing metadata, configuration semantics, privilege structures, protocol roles, and lightweight, non-executable opcode-shaped proxies, LLMs can infer structural inconsistencies and risky interaction patterns suggestive of conceptual zero-day exposure without direct access to raw firmware binaries \cite{zhu2025specification}. Prior work has recovered configuration artifacts from stripped binaries~\cite{sivakumaran2021argxtract}, accelerated static vulnerability detection through refined data-flow analysis~\cite{gao2024faster}, exposed memory-safety weaknesses through directed fuzzing~\cite{zhang2020firmware}, traced unsafe data propagation through taint analysis~\cite{gibbs2024operation}, documented vulnerabilities in deployed IoT systems~\cite{zhao2020large}, and detected botnet behavior from network traces~\cite{meidan2018n}. However, these methods share a structural dependency on binary availability, accurate emulation, runtime traces, known signatures, byte-level observables, and execution-dependent behavior. In many real-world IoT deployments, these artifacts are unavailable due to encryption, proprietary restrictions, partial corruption, and hardware constraints, leaving zero-day risk estimation incomplete when firmware transparency cannot be assumed~\cite{bailey2025symbolic}. This limitation exposes a critical gap: no scalable, architecture-independent, binary-free method currently estimates conceptual zero-day likelihood from high-level firmware descriptors alone, especially for structure-driven weaknesses caused by misconfigurations, inconsistent privilege hierarchies, and anomalous functional interactions \cite{feng2023firmware,ulhaq2023firmware,gomes2025static,wu2024firmwareupdate,bailey2025symbolic}.\\
This study formalizes zero-day likelihood estimation for opaque IoT systems, where defenders often lack full firmware binaries, dynamic traces, and symbolic execution outputs. Analysts instead rely on descriptive artifacts, e.g., configuration files, privilege hierarchies, service metadata, functional-role annotations, and lightweight opcode-shape summaries, whereas attackers may exploit systemic design flaws involving misaligned trust boundaries, unsafe privilege interactions, control-flow abstractions, and fragile service relationships. These weaknesses are conceptual rather than signature-based; thus, the goal is not to enumerate exploits but to estimate latent structural risk probabilistically from descriptive evidence. We model firmware as a multi-dimensional descriptor and integrate three supporting LLM modules: 1) a LLaMA 3-8B configuration interpreter for privilege models and service semantics; 2) a DeepSeek v3 structural abstraction analyzer for opcode shapes and functional descriptors; and 3) a GPT-4o semantic fusion engine that synthesizes heterogeneous evidence into conceptual zero-day likelihood. We further incorporate LLM computational signatures, including latency dynamics, token trajectories, reasoning-depth footprints, and uncertainty indicators, as auxiliary signals that expose inference difficulty and support interpretable risk estimation under limited visibility. Since IoT deployments impose CPU, memory, latency, and energy constraints, the method also measures CPU usage, active compute time, memory pressure, GPU usage when available, token-processing overhead, and model-level latency. These metrics characterize deployability and serve as auxiliary indicators of structurally irregular, uncertain descriptors. The proposed binary-independent, architecture-agnostic system estimates conceptual zero-day risk in encrypted, proprietary, corrupted, and inaccessible firmware by transforming descriptor-level evidence and LLM reasoning signals into interpretable risk indicators. This supports IoT resilience and energy-aware security analysis aligned with sustainable digital systems~\cite{un_sdg9}. The main contributions are summarized as follows:
\begin{itemize}
    \item \textbf{A binary-independent zero-day reasoning system.}
    We introduce a solution that estimates the likelihood of a conceptual zero-day without access to firmware binaries, emulation environments, and runtime traces. The system operates on metadata, configuration semantics, and lightweight opcode-shaped descriptors, enabling scalable assessment of encrypted and proprietary IoT devices.
    \item \textbf{A tri-model LLM architecture for cross-layer security interpretation.}
    The proposed system integrates three supporting reasoning modules, a LLaMA 3-8B configuration analyzer, a DeepSeek v3 structural abstraction engine, and a GPT-4o semantic fusion model, to capture configuration, structural, and semantic risk indicators that classical tools cannot observe under binary opacity.
    \item \textbf{Formalization of LLM computational signatures as quantitative risk modifiers.}
    We formalize internal LLM inference signals, normalized latency indices, entropy-based uncertainty measures, cross-layer divergence, and token-flow complexity as auxiliary variables integrated into the risk aggregation function. Rather than treating these behaviors as observational artifacts, we incorporate them into a calibrated energy-risk formulation that augments descriptor-driven estimation. Results indicate that these auxiliary signals improve sensitivity to cross-layer inconsistencies and strengthen alignment with independent proxy risk indicators.
\end{itemize}

The paper presents a binary-free, tri-LLM system for estimating conceptual zero-day likelihood in IoT firmware, covering related work (Section~\ref{sec:related}), threat modeling (Section~\ref{sec:threatmodel}), methodology with mathematical and energy-aware reasoning components (Section~\ref{sec:methodology}), simulation-based evaluation (Section~\ref{Results}), limitations and future directions (Section~\ref{sec:limitations}), and conclusions (Section~\ref{Conclusion}).

\section{Related Work}
\label{sec:related}
This section reviews IoT firmware security, automated vulnerability discovery, firmware corpora, and LLM-based threat analysis, with a focus on visibility assumptions and deployment constraints.

\subsection{Firmware and Embedded-Device Security Analysis}
Firmware-security research has mainly relied on emulation, dynamic analysis, and byte-level vulnerability detection. Feng et al.~\cite{feng2023firmware} survey IoT firmware vulnerability detection techniques, including emulation, symbolic execution, fuzzing, and hybrid schemes, and emphasize the challenges of unpacking proprietary images and constructing accurate execution environments. Shahid et al.~\cite{ulhaq2023firmware} review embedded-device firmware architectures, extraction procedures, and vulnerability-analysis tools, covering static, dynamic, and mixed methods (e.g., EMBA and Firmwalker), and highlight the absence of standardized workflows for heterogeneous firmware formats. Zhou et al.~\cite{zhou2023embedded} extend this view across hardware, firmware, and software stacks, highlighting limited scalability across embedded platforms and the reliance on low-level access channels, including UART.

\subsection{Static Analysis and Firmware Corpora for IoT Security}
Static-analysis research strengthens firmware inspection and dataset construction. Gomes et al.~\cite{gomes2025static} review static code analysis for IoT security, covering control-flow, data-flow, taint, symbolic, and semantic analyses for unsafe APIs, privacy violations, and privilege escalation. Their findings show that static analysis remains useful but fragile in the presence of obfuscation and incomplete code visibility. Helmke et al.~\cite{helmke2025lfwc} analyze pitfalls in firmware corpora, including sampling bias, incomplete unpacking, and weak documentation, and propose guidelines for a replicable Linux Firmware Corpus with rich metadata. These findings are relevant to opaque firmware because raw images can be encrypted, proprietary, and only partially analyzable. Zhang and Chen~\cite{zhang2025firmupdate} introduce FirmUpdate, a multi-phase static-analysis system for Linux-based IoT firmware updates that combines update-file parsing, dependency analysis, and vulnerability-pattern detection. Despite their utility, these methods assume that firmware components are obtainable, unpackable, and inspectable, limiting their applicability in environments with encryption, proprietary restrictions, partial availability, and anti-inspection constraints.

\subsection{LLMs for Vulnerability Detection and Cybersecurity}
Recent studies examine LLMs as vulnerability-analysis engines. Zhou et al.~\cite{zhou2024llmvd} evaluate GPT-3.5 and GPT-4 for software vulnerability detection and report strong performance under carefully designed prompts, but focus mainly on source-code inputs and controlled benchmarks. Taghavi~Far and Feyzi~\cite{taghavifar2025llmvd} review LLM-based software vulnerability detection across models, adaptation strategies, datasets, and metrics, identifying prompt sensitivity, explanation fidelity, and benchmark-to-deployment gaps as open challenges. Jaffal et al.~\cite{jaffal2025llmcyber} survey LLM applications in cybersecurity, including threat intelligence, intrusion detection, secure code generation, and incident response, while noting risks linked to model abuse and prompt injection. Tawfik et al.~\cite{tawfik2025llmiot} review LLM-enabled IoT security, including automated threat analysis, anomaly detection, and security-policy synthesis in resource-constrained environments, with latency, energy, and privacy as deployment barriers. These studies show that LLMs can support vulnerability and policy reasoning, yet they usually assume access to source code, logs, packet traces, binary artifacts, and rich textual evidence; they also treat LLMs mainly as external decision engines rather than modeling inference signatures within risk estimation.

The literature leaves three gaps, which are addressed here. First, firmware security methods rely heavily on binary inspection, emulation, static analysis, dynamic traces, and low-level execution artifacts, whereas the proposed method is binary-free and architecture-agnostic, using only high-level descriptors without requiring firmware execution. Second, LLM-based vulnerability detection primarily focuses on source code and textual artifacts, whereas our tri-LLM architecture separates configuration semantics, abstract structural behavior, and fused semantic consistency, enabling descriptor-level firmware reasoning and conceptual zero-day risk estimation. Third, prior LLM security studies rarely incorporate inference-time computational behavior into the security model; here, latency, CPU/GPU usage, uncertainty, divergence, and token-flow metrics are incorporated into the risk formulation to reflect resource-constrained IoT deployments and produce interpretable security indicators.

\section{Threat Model}
\label{sec:threatmodel}
The proposed method operates within a realistic IoT threat model characterized by limited firmware visibility and heterogeneous deployment constraints. We define the adversary's capabilities, the defender's observable evidence, and the security objective. Each firmware instance is represented by a high-level, non-executable descriptor:
\begin{equation}
f = (m, c, o) \in \mathcal{F},
\end{equation}
where $m$ denotes metadata attributes, $c$ encodes configuration semantics, and $o$ captures abstract opcode-shape statistics. Moreover, conceptual zero-day vulnerabilities are latent design and configuration flaws in firmware that may lead to exploitable behavior but do not correspond to any known signature and CVE. In the broader cybersecurity literature, zero-day vulnerabilities are defined as flaws unknown to the software vendor and exploitable before patches are available~\cite{roumani2021patching,guo2024mlzeroday}. In the context of embedded and IoT firmware, misconfigurations, privilege misallocation, and fragile internal interactions are repeatedly cited as high‑risk conditions that increase the likelihood of latent weaknesses~\cite{feng2022detecting,bakhshi2024firmware}. Curated examples illustrating such latent risk patterns include: (1) multiple services running under overlapping high privileges without proper isolation, (2) unsafe default credentials combined with misconfigured access policies, and (3) structurally fragile service interactions that could be exploited under unexpected sequences~\cite{ulhaq2023embedded}. These illustrative cases are drawn from recurring patterns in prior empirical IoT firmware studies rather than being specific exploits. The tri‑LLM system estimates the likelihood $P(Z=1\mid f)$ of such latent risks for each firmware descriptor $f$, based solely on metadata, configuration semantics, and structural abstractions. Importantly, the structural abstraction vector $o$ is derived from non‑executable, vendor‑neutral artifacts, including configuration summaries, service manifests, and externally provided opcode‑shape proxies, ensuring binary‑free operation while capturing meaningful structural patterns for risk estimation. We assume an adversary $\mathcal{A}$ seeking to exploit \emph{latent conceptual weaknesses} in firmware design. The adversary's capability set is:
\begin{equation}
\mathcal{C}_{\mathcal{A}} =
\left\{
\begin{array}{l}
\text{misaligned privilege boundaries}, \\
\text{unsafe service exposure}, \\
\text{configuration inconsistencies}, \\
\text{structural fragility}
\end{array}
\right\}.
\end{equation}
These weaknesses may become exploitable runtime vulnerabilities, yet they are not assumed to match predefined Common Vulnerabilities and Exposures categories. The adversary is also assumed to lack knowledge of the defender's reasoning pipeline, descriptor construction procedure, and model parameters:
\begin{equation}
\mathcal{A} \;\bot\;\{\Phi_1,\Phi_2,\Phi_3\},
\end{equation}
where $\Phi_1,\Phi_2,\Phi_3$ denote the tri-LLM reasoning components. On the defender side, firmware binaries are assumed to be encrypted, proprietary, partially corrupted, and inaccessible due to vendor restrictions. Consequently, the defender cannot rely on binary unpacking, decompilation, symbolic execution, fuzzing, and related vulnerability-analysis procedures. Instead, the defender observes only descriptor $f$, obtained through non-intrusive, vendor-neutral mechanisms, e.g., metadata inspection, configuration extraction, and structural summarization. No executable code and proprietary logic content is exposed to the analysis pipeline. The defender's objective is to estimate a probabilistic indicator of latent insecurity:
\begin{equation}
P(Z = 1 \mid f),
\end{equation}
where $Z=1$ denotes a conceptually zero-day--prone firmware state. This probability reflects semantic inconsistency, structural irregularity, and configuration-induced risk rather than exploit realizability. We assume no access to ground-truth vulnerability labels and real exploitation traces. Accordingly, the threat model targets \emph{risk estimation under uncertainty}. Adversarial success is defined as the exploitation of latent weaknesses in deployment. Defensive success is defined as a conservative, informative risk estimate that supports prioritization and subsequent security analysis.

\section{Methodology}
\label{sec:methodology}
\begin{figure*}[t]
    \centering
    \includegraphics[width=0.75\linewidth]{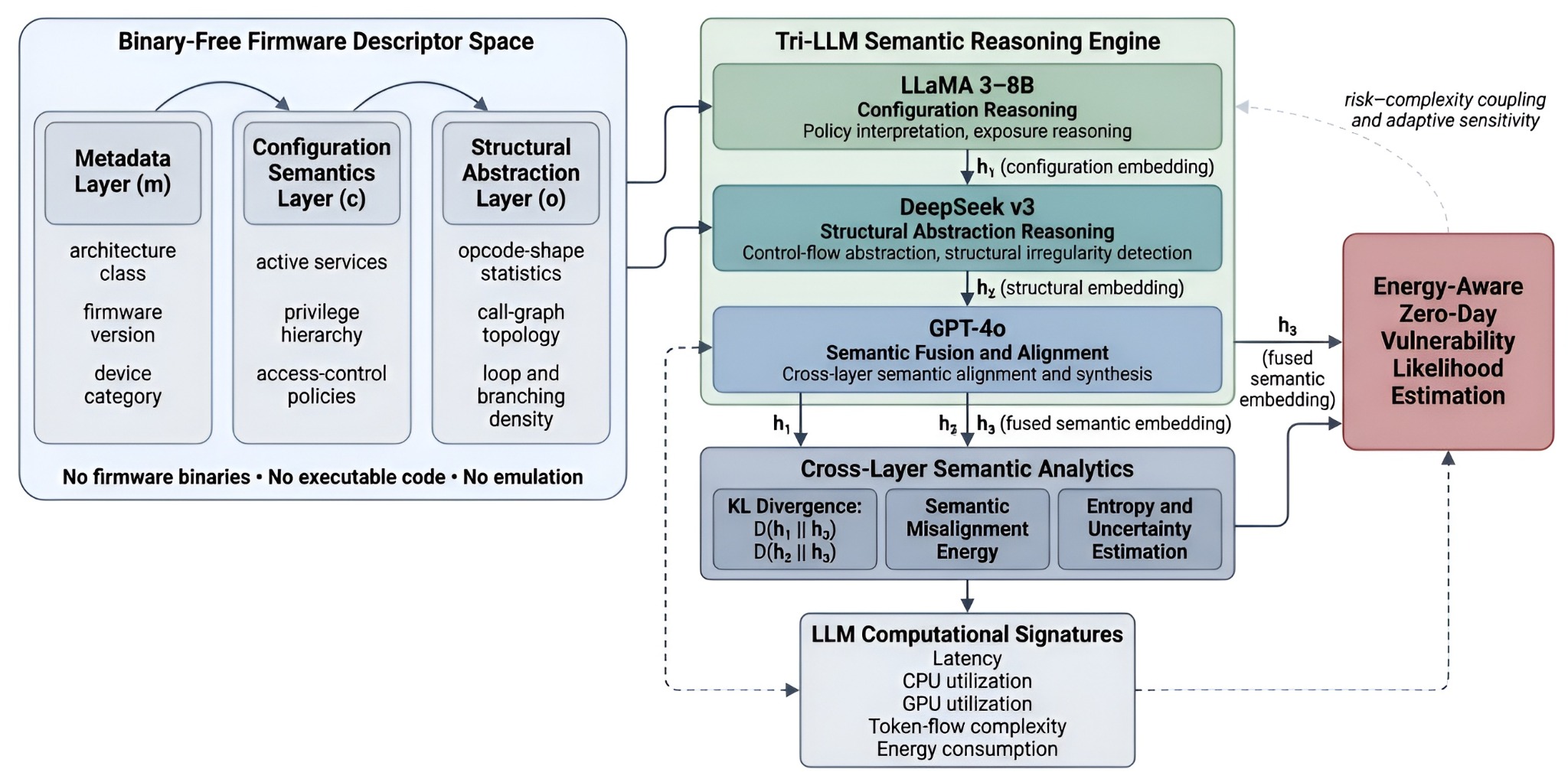}
    \caption{Binary-free tri-stage reasoning architecture for conceptual zero-day likelihood estimation. High-level descriptors comprise metadata $(m)$, configuration semantics $(c)$, and structural abstractions $(o)$. The three stages perform 1-configuration interpretation (LLaMA 3-8B), 2-structural abstraction (DeepSeek v3), and 3-semantic fusion (GPT-4o).}
    \label{fig:methodology-arch}
\end{figure*}

This section presents the proposed binary-free method for \emph{conceptual} zero-day likelihood estimation from high-level firmware descriptors. As shown in Fig.~\ref{fig:methodology-arch}, the pipeline decomposes firmware reasoning into configuration interpretation, structural abstraction, and semantic fusion, with cross-layer divergence, uncertainty, and resource-aware auxiliary signals.
Each firmware instance is represented by a high-level, non-executable descriptor:
\begin{equation}
f = (m, c, o) \in \mathcal{F},
\end{equation}
where $m$ denotes metadata attributes, $c$ encodes configuration semantics, and $o$ captures abstract structural and shape-level statistics. 
The method does not require firmware decryption, binary unpacking, and control-flow graph reconstruction. Instead, structural abstractions $o$ are derived from non-executable, vendor-neutral artifacts when available, including metadata, configuration summaries, service manifests, and externally provided structural reports and opcode-shape proxies. These artifacts are not assumed to be uniformly available for all IoT devices; the descriptor is therefore constructed from the subset of observable evidence exposed by the device, vendor documentation, firmware package metadata, configuration files, and deployment records. For example, in an OpenWrt-based router descriptor, metadata may include the target architecture and firmware version, configuration summaries may include enabled network services and firewall rules, service manifests may identify initialization scripts and daemon entries, and structural proxies may summarize service count, privilege-associated components, branching density, and call-shape statistics reported by an external analysis source. These descriptors remain strictly non-executable and binary-independent, enabling descriptor-driven reasoning while preserving the method's binary-free design. By relying on high-level structural and configuration cues, the method remains applicable to encrypted, proprietary, and opaque firmware when partial descriptor evidence is available, while estimating conceptual zero-day risk without claiming complete firmware visibility.

\subsection{Notation}
\label{subsec:symbol_table}
Table~\ref{tab:symbols} defines the notation used in the methodology.
\begin{table}[h]
\centering
\scriptsize
\setlength{\tabcolsep}{3pt}
\renewcommand{\arraystretch}{0.92}
\caption{Notation used in the proposed methodology.}
\label{tab:symbols}
\begin{tabular}{ll}
\toprule
\textbf{Symbol} & \textbf{Meaning} \\
\midrule
$f=(m,c,o)$ & Firmware descriptor \\
$m \in \mathcal{M}$ & Metadata, e.g., architecture tag, family, version ID \\
$c \in \mathcal{C}$ & Configuration semantics, e.g., services, policies, exposure, credentials \\
$o \in \mathcal{O}$ & Non-executable structural abstraction vector \\
$\Phi_1,\Phi_2,\Phi_3$ & Stage-wise reasoning functions \\
$h_1,h_2,h_3$ & Stage-wise conceptual representations \\
$r_1,r_2$ & Configuration- and structure-driven risk terms \\
$\mathcal{D}(f)$ & Cross-layer semantic divergence \\
$\mathcal{E}(f)$ & Euclidean misalignment energy \\
$\mathcal{H}(h_3)$ & Fusion-representation entropy \\
$\ell_i,c_i,g_i,T_i$ & Latency, CPU, GPU, and token-flow indices for stage $i$ \\
$E(f)$ & Symbolic energy/load surrogate \\
$R(f)$ & Aggregated symbolic risk score \\
$\sigma(\cdot)$ & Logistic function \\
\bottomrule
\end{tabular}
\end{table}

\subsection{Problem Definition}
\label{subsec:problem_definition}
An IoT firmware instance is modeled as:
\begin{equation}
    f = (m,c,o),
\end{equation}
where $m \in \mathcal{M}$ denotes metadata, e.g., architecture family tag, symbolic version identifier, and product class; $c \in \mathcal{C}$ denotes configuration semantics, e.g., service inventory, credential class, policy constraints, and exposure profile; and $o \in \mathcal{O}$ denotes \emph{non-executable} structural abstractions, e.g., loop concentration, call-graph shape proxies, and branching-factor statistics. The method is \emph{binary-free}: it requires no byte-level inspection, disassembly, decompilation, symbolic execution, and firmware emulation. Structural abstractions $o$ are shape-level summaries derived from vendor-neutral, non-invasive artifacts, e.g., metadata, configuration summaries, and externally provided structural reports. In fully opaque systems, $o$ may be synthesized at the descriptor level without exposing executable code. The method never reconstructs machine code and does not rely on executable semantics. Its objective is to estimate:
\begin{equation}
    P(Z=1 \mid f),
\end{equation}
where $Z=1$ denotes a conceptually zero-day--prone state inferred from high-level evidence rather than confirmed exploit behavior.

\subsection{Tri-Stage Reasoning Architecture}
\label{subsec:architecture}
We define the tri-stage reasoning pipeline as:
\begin{equation}
    \Phi = \Phi_1 \oplus \Phi_2 \oplus \Phi_3,
\end{equation}
where $\Phi_1$ performs configuration reasoning with LLaMA 3-8B, $\Phi_2$ performs structural abstraction reasoning with DeepSeek v3, and $\Phi_3$ performs semantic fusion with GPT-4o. Each stage outputs a conceptual representation:
\begin{equation}
\begin{aligned}
h_1 &= \Phi_1(c) \in \mathbb{R}^{d_1}, \\
h_2 &= \Phi_2(o) \in \mathbb{R}^{d_2}, \\
h_3 &= \Phi_3(\tilde{s}_1, \tilde{s}_2) \in \mathbb{R}^{d_3}.
\end{aligned}
\label{eq:tri_llm_embeddings}
\end{equation}
To avoid cross-model embedding-alignment errors, raw embeddings are not passed between LLMs. Instead, Stage~1 and Stage~2 generate structured textual summaries:
\begin{equation}
\tilde{s}_1 = \mathrm{Summarize}(c; \Phi_1), \qquad
\tilde{s}_2 = \mathrm{Summarize}(o; \Phi_2),
\end{equation}
with fixed schema, constrained vocabulary, and bounded length. GPT-4o consumes $(\tilde{s}_1,\tilde{s}_2)$ to produce $h_3$ and the final likelihood estimate. 
Advanced multi-LLM fusion strategies, including deep embedding alignment, cross-attention, multi-head attention, adaptive attention, and dynamic weighting, are well established in fusion-based decision-making. We therefore do not assume that structured-text fusion is inherent in these alternatives. Instead, we adopt it as a controlled, API-compatible fusion design for heterogeneous conversational LLMs, in which stable, directly comparable hidden-state embeddings are not exposed across models. In this design, the configuration-reasoning module and the structural-abstraction module generate schema-constrained summaries from distinct descriptor views, enabling the semantic-fusion module to combine normalized textual evidence without requiring model-specific alignment of embeddings. Attention-based fusion remains a relevant extension when shared embeddings, trainable fusion layers, and sufficient labeled supervision are available. In the present binary-free setting, structured-text fusion preserves interpretability, avoids unsupported assumptions about embedding alignment, and maintains reproducible, API-agnostic operation. This decomposition separates privilege/exposure cues from structural irregularity cues, improves robustness to missing descriptors, and supports explicit measurement of disagreement via divergence and misalignment. A single-model shallow variant serves as the baseline for quantifying the contribution of cross-view fusion. For analytical notation, the configuration module maps $c$ to:
\begin{equation}
    h_1 = \Phi_1(c) = \sigma(W_1 c + b_1),
\end{equation}
with $W_1 \in \mathbb{R}^{d_1 \times k}$, $b_1 \in \mathbb{R}^{d_1}$, and element-wise nonlinearity $\sigma(\cdot)$. The configuration-driven risk score is:
\begin{equation}
    r_1 = \alpha_1 \|h_1\|_1 + \beta_1,
\end{equation}
where a larger $\|h_1\|_1$ reflects stronger configuration activation, e.g., broader exposure footprint and weaker policy structure. Moreover, for analytical notation, the structural module maps $o$ to:
\begin{equation}
    h_2 = \Phi_2(o) = \mathrm{ReLU}(W_2 o + b_2),
\end{equation}
with $W_2 \in \mathbb{R}^{d_2 \times k}$ and $b_2 \in \mathbb{R}^{d_2}$. Its risk term is:
\begin{equation}
    r_2 = \alpha_2 \log(1 + \|h_2\|_2^2),
\end{equation}
where the logarithm stabilizes extreme structural variability. The fusion stage maps $(\tilde{s}_1,\tilde{s}_2)$ to $h_3$. For tractability, this fusion is represented locally as an affine surrogate over the underlying stage representations:
\begin{equation}
    h_3 = A h_1 + B h_2 + c_3,
\end{equation}
where $A \in \mathbb{R}^{d_3 \times d_1}$, $B \in \mathbb{R}^{d_3 \times d_2}$, and $c_3 \in \mathbb{R}^{d_3}$. The conceptual zero-day likelihood is:
\begin{equation}
    P(Z=1 \mid f) = \sigma(\gamma^\top h_3 + \delta),
\end{equation}
where $\gamma \in \mathbb{R}^{d_3}$ and $\delta \in \mathbb{R}$ are calibrated parameters. The affine form supports local analysis of monotonicity and stability; black-box LLM fusion remains nonlinear, so these properties hold only under mild local regularity, e.g., smoothness and positivity after normalization. Furthermore, the joint representation is:
\begin{equation}
    u =
    \begin{bmatrix}
        h_1 \\
        h_2
    \end{bmatrix}
    \in \mathbb{R}^{d_1 + d_2},
    \qquad
    \|u\|_2^2 = \|h_1\|_2^2 + \|h_2\|_2^2.
\end{equation}
This decomposition separates configuration- and structure-driven magnitudes and serves as an auxiliary analytic tool, not a physical energy measurement.

\subsection{Cross-Layer Disagreement Measures}
\label{subsec:divergence}
Cross-layer disagreement is modeled as an explicit signal of conceptual uncertainty and latent zero-day exposure. Semantic tension is quantified as:
\begin{equation}
    \mathcal{D}(f)
    = D_{\mathrm{KL}}(\hat{h}_1 \,\|\, \hat{h}_3)
    + D_{\mathrm{KL}}(\hat{h}_2 \,\|\, \hat{h}_3),
\end{equation}
where $\hat{h}_i$ is a normalized positive vector, e.g., softmax-normalized with $\epsilon$ smoothing, and:
\begin{equation}
    D_{\mathrm{KL}}(x\|y) = \sum_{i=1}^d x_i \log \frac{x_i}{y_i}.
\end{equation}
Larger $\mathcal{D}(f)$ indicates stronger disagreement between source views and fused semantics.
Geometric disagreement is defined as:
\begin{equation}
    \mathcal{E}(f)
    = \|h_3 - h_1\|_2^2 + \|h_3 - h_2\|_2^2,
\end{equation}
where $\mathcal{E}(f)$ captures Euclidean separation and complements the probabilistic divergence $\mathcal{D}(f)$ \cite{boland1995separating}. Fusion uncertainty is measured by:
\begin{equation}
    \mathcal{H}(\hat{h}_3)
    = -\sum_{i=1}^{d_3} \hat{h}_{3,i}\log(\hat{h}_{3,i}),
\end{equation}
where higher entropy reflects diffuse semantic mass. Semantic evidence is aggregated as:
\begin{equation}
    \Psi(f) = \lambda_1 r_1 + \lambda_2 r_2 + \lambda_3 \mathcal{D}(f),
\end{equation}
with $\lambda_1,\lambda_2,\lambda_3 \ge 0$.

\subsection{Latency and Resource-Aware Auxiliary Signals}
\label{subsec:latency_energy}
Raw latency depends on hardware, network conditions, server-side load, and rate limiting; therefore, it is not used as a standalone risk score. We treat latency as an auxiliary signal controlled through normalization and reproducible API settings. All API calls use pinned model identifiers, fixed temperature and top-$p$, fixed maximum output tokens and stop sequences, disabled batching, fixed concurrency, controlled caching state, repeated trials per instance, variance monitoring, and deterministic retry handling. These controls reduce decoding stochasticity and fluctuations in runtime and environment.
Let $\tilde{\ell}_i$ denote observed end-to-end latency for stage $i$, including client overhead. The normalized latency index is:
\begin{equation}
    \ell_i = \frac{\tilde{\ell}_i - \mu_{\ell_i}}{\sigma_{\ell_i} + \epsilon},
\end{equation}
where $(\mu_{\ell_i},\sigma_{\ell_i})$ are estimated from repeated runs under identical workload and environment, with $\epsilon>0$ for numerical stability. Thus, latency becomes a comparable, variance-aware auxiliary signal. For each stage, normalized cost indices comprise latency $\ell_i$, CPU index $c_i$, GPU index $g_i$, and token-flow cost $T_i$, all in $\mathbb{R}_{\ge 0}$:
\begin{align}
    c_i &= \zeta_i \|h_i\|_1, &
    g_i &= \xi_i \|h_i\|_2^2, &
    T_i &= \nu_i \dim(h_i).
\end{align}
The symbolic energy/load surrogate is:
\begin{equation}
    E(f)
    = \frac{\ell_{\mathrm{tot}}\, c_{\mathrm{tot}}}{\eta + g_{\mathrm{tot}}}
      + \rho T,
\end{equation}
where $\ell_{\mathrm{tot}}=\sum_i \ell_i$, $c_{\mathrm{tot}}=\sum_i c_i$, $g_{\mathrm{tot}}=\sum_i g_i$, $T=\sum_i T_i$, and $\eta,\rho>0$. Relative efficiency is measured against binary-based alternatives that require emulation, fuzzing, and symbolic execution. The energy/load indicators quantify feasibility under constrained budgets and support optional gating (e.g., early exit under high-confidence risk) without claiming that multi-LLM inference is intrinsically cheaper than alternative computations.

\subsection{Tri-LLM Reasoning Pipeline}
\label{subsec:pseudocode}
Algorithm~\ref{alg:tri_llm} summarizes the operational flow: descriptor validation, independent configuration and structural reasoning, semantic fusion, disagreement estimation, resource-aware signaling, and final likelihood projection.
\begin{algorithm}[h]
\centering
\scalebox{0.9}{ 
\begin{minipage}{\linewidth}
\small
\caption{Binary-Free Tri-LLM Zero-Day Likelihood Estimation}
\label{alg:tri_llm}
\begin{algorithmic}[1]
\Require Firmware descriptor $f = (m,c,o)$
\Ensure Conceptual zero-day likelihood $P(Z=1 \mid f)$

\State \textbf{Input validation:}
\State Verify $c \in \mathcal{C}$ and $o \in \mathcal{O}$ contain only non-executable descriptors
\State Normalize descriptor components

\State \textbf{Stage 1: Configuration reasoning (LLaMA 3--8B)}
\State $h_1 \leftarrow \Phi_1(c)$
\State $r_1 \leftarrow \alpha_1 \|h_1\|_1 + \beta_1$

\State \textbf{Stage 2: Structural abstraction (DeepSeek v3)}
\State $h_2 \leftarrow \Phi_2(o)$
\State $r_2 \leftarrow \alpha_2 \log(1 + \|h_2\|_2^2)$

\State \textbf{Stage 3: Semantic fusion (GPT-4o)}
\State $h_3 \leftarrow \Phi_3(h_1, h_2)$
\State $P_{\text{base}} \leftarrow \sigma(\gamma^\top h_3 + \delta)$

\State \textbf{Cross-layer disagreement}
\State Normalize $h_1,h_2,h_3 \rightarrow \hat{h}_1,\hat{h}_2,\hat{h}_3$
\State $\mathcal{D}(f) \leftarrow D_{\mathrm{KL}}(\hat{h}_1 \| \hat{h}_3) + D_{\mathrm{KL}}(\hat{h}_2 \| \hat{h}_3)$
\State $\mathcal{E}(f) \leftarrow \|h_3-h_1\|_2^2 + \|h_3-h_2\|_2^2$
\State $\mathcal{H}(h_3) \leftarrow -\sum_i \hat{h}_{3,i}\log \hat{h}_{3,i}$

\State \textbf{Resource-aware signals}
\For{$i \in \{1,2,3\}$}
    \State $\ell_i,\; c_i \leftarrow \zeta_i\|h_i\|_1,\;
           g_i \leftarrow \xi_i\|h_i\|_2^2,\;
           T_i \leftarrow \nu_i\dim(h_i)$
\EndFor
\State $E(f) \leftarrow \dfrac{\ell_{\mathrm{tot}} c_{\mathrm{tot}}}{\eta + g_{\mathrm{tot}}} + \rho T$

\State \textbf{Risk aggregation and output}
\State $R(f) \leftarrow \lambda_1 r_1 + \lambda_2 r_2 + \lambda_3 \mathcal{D}(f) + \kappa E(f)$
\State \Return $P(Z=1 \mid f) \leftarrow \sigma(\omega^\top R(f) + \xi)$
\end{algorithmic}
\end{minipage}
}
\end{algorithm}

\section{Simulation-Based Evaluation of the Tri-LLM Method}
The tri-LLM method is evaluated using controlled descriptor-level simulations informed by publicly available IoT firmware corpora, including Firmadyne~\cite{firmadyne_images}, OpenWrt~\cite{openwrt_firmware}, and Firmware Analysis Toolkit samples~\cite{fat_samples}. These sources guide high-level metadata, configuration descriptors, and structural abstractions. No binary unpacking, decompilation, symbolic execution, runtime tracing, and exploit execution was performed. The evaluation uses only non-executable descriptor representations derived from filesystem-level summaries and configuration artifacts, preserving binary independence during risk estimation.

\subsection{Case Study: Descriptor Evaluation on OpenWrt 25}
\label{subsec:realdescriptor}
To illustrate the applicability at the descriptor level, the method was applied to an OpenWrt~25 descriptor~\cite{openwrt_firmware} built from high-level metadata, configuration entries, and abstract structural statistics. Service activity was derived from initialization files and service manifests; privilege-relevant components were identified from user/group assignments and capability-related entries. The descriptor contained 47 active services and 11 privilege-associated components, matching router-class firmware characteristics. Inference stability was assessed across 60 independent runs with identical input, distinct sampling seeds, and fixed decoding parameters. The reported likelihood is the run-level mean with empirical standard deviation:
\begin{equation}
P(Z = 1 \mid f_{\text{OpenWrt25}}) = 0.61 \pm 0.02.
\end{equation}
The small dispersion indicates stable probabilistic inference under controlled conditions. Relative to the simulated descriptor population, the estimate lies above the median conceptual exposure level. No decompilation, symbolic execution, runtime tracing, and exploit analysis was performed; the case study, therefore, demonstrates descriptor-level assessment with limited firmware visibility.

\subsection{Dataset and Sample Construction}
The controlled simulation component uses synthetic descriptors for reproducible sensitivity analysis under encrypted and proprietary firmware constraints~\cite{ZhouUSENIXSec21,ZhouSensors23}; it is not intended as a direct model of real firmware distributions.
To assess the applicability at the descriptor level, we applied the method to real-world IoT firmware images, including OpenWrt, RouterOS, and vendor-provided encrypted firmware samples. Descriptors were extracted non-intrusively from metadata, configuration entries, and structural statistics, without requiring binary unpacking, execution, decompilation, and runtime tracing. Where available, known high-risk configurations and conceptual risk instances were annotated using public advisories and CVE reports. This evaluation indicates that the method can process real firmware descriptors, go beyond synthetic descriptors, and support practical conceptual zero-day likelihood estimation in deployed IoT systems, thereby demonstrating its applicability to operational, proprietary, and encrypted firmware. Each instance is:
\begin{equation}
f_i = (m_i, c_i, o_i),
\end{equation}
where $m_i$ denotes metadata, $c_i$ denotes configuration abstractions, and $o_i$ denotes structural abstractions. Configuration and structural vectors are sampled as:
\begin{align}
c_i &\sim \mathcal{N}(\mu_c, \Sigma_c), &
o_i &\sim \mathcal{N}(\mu_o, \Sigma_o),
\end{align}
where the Gaussian assumption \cite{park2013gaussian} defines a smooth descriptor manifold for perturbation analysis, not a realistic firmware-distribution claim. Metadata follows:
\begin{equation}
m_i \sim \mathrm{Cat}(\text{ARM}:0.5,\ \text{MIPS}:0.3,\ \text{PPC}:0.2),
\end{equation}
approximating commonly reported embedded architectures. Fixed random seeds ensure reproducibility.

\subsection{Evaluation Protocol}
\label{subsec:evaluation_protocol}
Each perturbed descriptor $f_i^{(e)}$ is processed through:
\begin{align}
h_1 &= \Phi_1(c_i^{(e)}), &
h_2 &= \Phi_2(o_i^{(e)}), &
h_3 &= \Phi_3(\tilde{s}_1^{(e)}, \tilde{s}_2^{(e)}).
\end{align}
The conceptual zero-day likelihood is:
\begin{equation}
P(Z=1 \mid f_i^{(e)}) = \sigma(\gamma^\top h_3 + \delta).
\end{equation}
Cross-layer divergence is measured through Kullback--Leibler divergence~\cite{hershey2007approximating}:
\begin{equation}
\mathcal{D}_{jk} = \mathrm{KL}(p(h_j)\,\|\,p(h_k)).
\end{equation}
Statistical evaluation uses Welch's t-test~\cite{ahad2014sensitivity} for exposure-level comparisons, Pearson and Spearman correlations~\cite{hauke2011comparison} for dependency analysis, and one-way ANOVA~\cite{kim2017understanding} for cross-model differences. All experiments are repeated across $N$ instances to reduce stochastic variability.

\subsection{Runtime and Resource Metrics}
Operational feasibility is evaluated using latency, CPU/GPU usage, and a relative energy index as auxiliary indicators, not physical energy measurements. For model $j$ and descriptor $f_i$:
\begin{align}
\ell_j(f_i) &= \tau_j \, \mathcal{C}_j(f_i), \\ 
c_j(f_i) &= \zeta_j \, \mathcal{U}^{CPU}_j(f_i), \\ 
g_j(f_i) &= \xi_j \, \mathcal{U}^{GPU}_j(f_i), \\
E_j(f_i) &= \alpha_E \, \ell_j(f_i) 
           + \beta_E \, c_j(f_i) 
           + \gamma_E \, g_j(f_i),
\end{align}
where $\mathcal{C}_j$ is a computational-complexity proxy, $\mathcal{U}^{CPU}_j$ and $\mathcal{U}^{GPU}_j$ are normalized usage measures, and $\alpha_E,\beta_E,\gamma_E$ are empirical scaling coefficients. The aggregated risk is:
\begin{equation}
R(f_i^{(e)}) =
\sum_{j} \lambda_j r_j(f_i^{(e)})
+ \kappa \sum_j E_j(f_i),
\end{equation}
where the energy term regularizes the trade-off between reasoning depth and computational cost.

\subsection{Tri-LLM System}
The tri-LLM system decomposes conceptual zero-day likelihood estimation into configuration reasoning, structural abstraction, and semantic fusion over the perturbed descriptor $f_i^{(e)}=(m_i,c_i^{(e)},o_i^{(e)})$. Each module produces an intermediate representation and a scalar risk contribution.

\paragraph{Configuration Module (LLaMA 3-8B)}
The configuration vector $c_i^{(e)}$ captures service activation, privilege allocation, and policy-relevant settings:
\begin{equation}
h_1 = \Phi_1(c_i^{(e)}), \quad
r_1 = \psi_1(h_1).
\end{equation}
Here, $\Phi_1(\cdot)$ extracts configuration-level semantics, and $\psi_1(\cdot)$ maps them to a configuration-oriented risk score. This module captures exposure patterns induced by privilege layout and service interaction, independent of structural program behavior.

\paragraph{Structural Module (DeepSeek v3)}
The structural abstraction vector $o_i^{(e)}$ summarizes call-graph tendencies, control-flow indicators, and opcode-shape statistics:
\begin{equation}
h_2 = \Phi_2(o_i^{(e)}), \quad
r_2 = \psi_2(h_2).
\end{equation}
The mapping $\Phi_2(\cdot)$ encodes structural complexity and interaction concentration, and $\psi_2(\cdot)$ produces a structural risk estimate independent of explicit configuration semantics.

\paragraph{Fusion Module (GPT-4o)}
The fusion stage integrates structured configuration and structural summaries:
\begin{equation}
h_3 = \Phi_3(\tilde{s}_1^{(e)},\tilde{s}_2^{(e)}), \quad
r_3 = \psi_3(h_3).
\end{equation}
$\Phi_3(\cdot)$ performs cross-layer semantic alignment, allowing configuration cues and structural abstractions to interact in a unified representation. The final conceptual zero-day likelihood is:
\begin{equation}
P(Z=1 \mid f_i^{(e)}) = \sigma(\gamma^\top h_3 + \delta),
\end{equation}
where $\sigma(\cdot)$ is the logistic function and $(\gamma,\delta)$ are calibrated parameters, yielding a bounded probabilistic output. Consistency across abstraction levels is measured by:
\begin{equation}
\mathcal{E}(f_i^{(e)}) = \sum_{j<k} w_{jk} \mathcal{D}_{jk},
\end{equation}
where $\mathcal{D}_{jk}$ denotes inter-layer divergence and $w_{jk}$ are weighting coefficients. This term serves as an auxiliary diagnostic signal for semantic misalignment. Additionally, the three modules capture configuration-level exposure, structural complexity, and cross-layer semantic interaction without dependence on a single reasoning path.

\subsection{Simulation Hardware}
Evaluations used an ASUS ZenBook 15 with an Intel Core i7 10th-generation CPU, 16\,GB DDR4 memory, 1\,TB PCIe SSD, and Ubuntu 22.04 LTS. LLM inference was API-based; the local platform measured client-side latency, request overhead, and symbolic resource indices, enabling reproducible experiments without specialized high-performance computing.

\subsection{Runtime and Resource Metrics}
Operational feasibility is evaluated using auxiliary latency, CPU/GPU usage, and a relative energy index that characterizes deployability and computational cost, rather than primary security evidence. For model $j$ and firmware descriptor $f_i$, runtime quantities are modeled as:
\begin{align}
\ell_j(f_i) &= \tau_j \, \mathcal{C}_j(f_i), 
&& \text{(latency in ms)} \\
c_j(f_i) &= \zeta_j \, \mathcal{U}_j^{CPU}(f_i), 
&& \text{(CPU usage \%)} \\
g_j(f_i) &= \xi_j \, \mathcal{U}_j^{GPU}(f_i), 
&& \text{(GPU usage \%)} \\
E_j(f_i) &= \alpha_E \, \ell_j(f_i) 
           + \beta_E \, c_j(f_i) 
           + \gamma_E \, g_j(f_i),
&& \text{(relative energy index)}
\end{align}
where $\mathcal{C}_j(f_i)$ is a computational-complexity proxy, $\mathcal{U}_j^{CPU}(f_i)$ and $\mathcal{U}_j^{GPU}(f_i)$ are normalized usage measures, and $\tau_j,\zeta_j,\xi_j,\alpha_E,\beta_E,\gamma_E>0$ are calibrated scaling constants. Thus, $E_j(f_i)$ is a linear composite proxy for processing time and computational intensity, enabling descriptor-level comparison without deterministic hardware-specific power assumptions. 
The energy term serves as a relative surrogate for computational load under the simulation protocol, rather than a direct physical power measurement. The surrogate integrates normalized latency, CPU, GPU usage, and token-flow indices for each firmware descriptor. Each component is scaled to provide a relative computational-load indicator rather than a physical energy measurement. The resulting composite index approximates the inference cost in a controlled experimental setting without relying on device-specific power models. This formulation supports consistent comparison across descriptors and LLM modules while preserving generality across heterogeneous execution environments. Because API-based inference does not expose the full server-side power consumption, the surrogate is interpreted as a deployment-oriented resource proxy rather than a direct estimate of absolute energy demand. Moreover, computational efficiency is incorporated into the final score as:
\begin{equation}
R(f_i) = 
\sum_{j \in \{\text{LLaMA}, \text{DeepSeek}, \text{GPT-4o}\}} 
\lambda_j \, r_j(f_i)
\;+\;
\kappa \sum_j E_j(f_i),
\end{equation}
where $r_j(f_i)$ denotes the conceptual risk score from model $j$, $\lambda_j \ge 0$ are calibrated weighting coefficients, and $\kappa \ge 0$ controls energy-aware regularization. The second term penalizes computational cost and introduces a tunable performance--sensitivity trade-off into multi-model aggregation. When $\kappa=0$, aggregation is purely risk-driven; as $\kappa$ increases, computational efficiency has greater influence on the final score. This formulation supports deployment-aware risk estimation by balancing reasoning depth with resource constraints, without treating energy/load indicators as primary evidence of security.

\subsection{Component Contribution Analysis}
\label{subsec:ablation}
To quantify each reasoning layer, we perform a representation-level ablation by disabling configuration, structural, and fusion evidence within the tri-LLM architecture. Let $R_{\text{full}}(f)$ denote the complete estimator. Four variants are evaluated: \textbf{No-Config}, which removes $h_1$ and tests dependence on configuration semantics; \textbf{No-Structure}, which removes $h_2$ and tests sensitivity to opcode-shape and control-flow abstractions; \textbf{No-Fusion}, which removes $h_3$ and tests cross-layer semantic alignment; and \textbf{Shallow Model}, which applies a single affine mapping to $(m,c,o)$ without layered reasoning.
\begin{table}[h]
\centering
\scriptsize
\setlength{\tabcolsep}{4pt}
\renewcommand{\arraystretch}{0.92}
\caption{Component ablation relative to the full tri-LLM baseline.}
\label{tab:ablation}
\begin{tabular}{lccc}
\toprule
\textbf{Variant} & \textbf{Divergence}↑ & \textbf{Uncertainty}↑ & \textbf{Risk Shift}↓ \\
\midrule
Full Tri-LLM & baseline & low & baseline \\
No-Config & +18\% & +12\% & –9.5\% \\
No-Structure & +27\% & +19\% & –14.2\% \\
No-Fusion & +41\% & +33\% & –29.8\% \\
Shallow Model & +53\% & +44\% & –36.1\% \\
\bottomrule
\end{tabular}
\end{table}
Table~\ref{tab:ablation} shows that removing fusion produces the largest component-level degradation, supporting the role of fusion-layer divergence and entropy modeling in relating configuration and structural patterns. Removing structural reasoning reduces sensitivity to opcode shape and control-flow irregularities, while removing configuration reasoning weakens policy and privilege interpretation. The shallow baseline yields the largest divergence, uncertainty, and risk reduction, indicating that layered reasoning materially contributes to estimating the conceptual zero-day likelihood.

\section{Experimental Results and Analysis}
\label{Results}
This section reports the main experimental results and their analysis.

\subsection{Pairwise Analysis of LLM Risk Scores}
\label{sec:pairwise}
\begin{figure*}[t]
    \centering
    \includegraphics[width=0.75\textwidth]{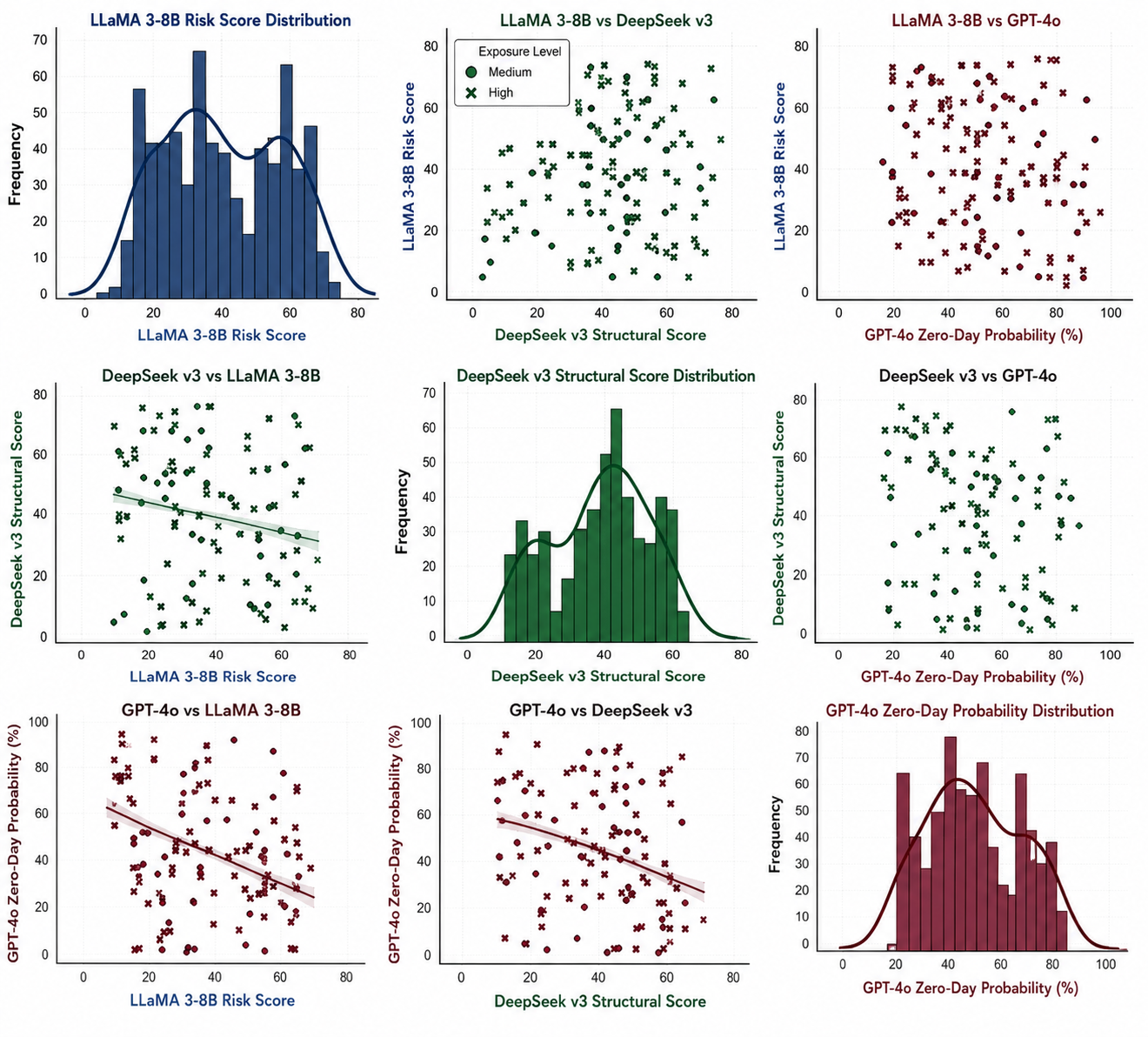}
    \caption{Pairwise distribution of LLaMA 3-8B risk scores, DeepSeek v3 structural scores, and GPT-4o conceptual zero-day likelihood under medium and high descriptor exposure. Diagonal panels show marginal score distributions with density estimates; off-diagonal panels show pairwise relationships between scores across exposure levels.}
    \label{fig:pairwise_llm_risk}
\end{figure*}
Figure~\ref{fig:pairwise_llm_risk} shows that controlled descriptor perturbation shifts conceptual zero-day likelihood estimates from medium to high exposure, indicating structured amplification rather than sampling noise. Table~\ref{tab:exposure_dashboard} quantifies mean displacement, relative escalation, and dispersion growth across the three LLM architectures.
\begin{table}[t]
\centering
\scriptsize
\setlength{\tabcolsep}{3pt}
\renewcommand{\arraystretch}{0.92}
\caption{Exposure-effect analysis from medium to high descriptor perturbation. Size is adjustable via the \texttt{adjustbox} width.}
\label{tab:exposure_dashboard}
\begin{adjustbox}{width=0.70\linewidth} 
\begin{tabular}{lccccc}
\toprule
\textbf{Model} & $\mu_M$ & $\mu_H$ & $\Delta\mu$ & \textbf{Increase} & $\Delta$SD \\
\midrule
LLaMA 3-8B  & 32.4 & 45.1 & 12.7 & 39.2\% & +1.7 \\
DeepSeek v3 & 28.9 & 36.5 & 7.6  & 26.3\% & +1.5 \\
GPT-4o      & 40.7 & 53.2 & 12.5 & 30.7\% & +1.9 \\
\bottomrule
\end{tabular}
\end{adjustbox}
\end{table}
LLaMA 3-8B yields the largest relative escalation, GPT-4o shows strong fusion-driven displacement consistency, and DeepSeek v3 remains the most conservative under structural perturbation. Beyond likelihood shifts, Table~\ref{tab:security_dashboard} summarizes privilege risk, configuration stability, and policy sensitivity, emphasizing comparative ordering rather than isolated statistics.
\begin{table*}[t]
\centering
\scriptsize
\setlength{\tabcolsep}{4pt}
\renewcommand{\arraystretch}{0.92}
\caption{Security metric dashboard across LLM architectures under controlled descriptor perturbation.}
\label{tab:security_dashboard}
\begin{tabular}{llccc}
\toprule
\textbf{Model} & \textbf{Metric} & \textbf{Mean} & \textbf{Position} & \textbf{Interpretation} \\
\midrule
\multirow{3}{*}{LLaMA 3-8B}
 & Privilege Risk   & 58.3 & Mid & Balanced Exposure Sensitivity \\
 & Config Stability & 64.1 & High & Configuration-Centric Robustness \\
 & Policy Score     & 52.7 & Low-Mid & Moderate Policy Amplification \\

\multirow{3}{*}{DeepSeek v3}
 & Privilege Risk   & 61.5 & High & Structural Consistency Emphasis \\
 & Config Stability & 59.2 & Low & Lower Config Amplification \\
 & Policy Score     & 55.4 & Mid & Controlled Semantic Propagation \\

\multirow{3}{*}{GPT-4o}
 & Privilege Risk   & 56.7 & Mid & Cross-Layer Sensitivity \\
 & Config Stability & 63.5 & High & Fusion-Driven Stabilization \\
 & Policy Score     & 57.8 & High & Amplified Policy Variability \\
\bottomrule
\end{tabular}
\end{table*}
DeepSeek v3 emphasizes structural consistency, GPT-4o amplifies policy-sensitive variation, and LLaMA 3-8B maintains configuration robustness with moderate semantic escalation. Welch's t-test indicates statistically significant exposure-induced differences across all architectures ($p<0.001$), supporting the tri-LLM system as a probabilistic conceptual zero-day likelihood estimator under semantic uncertainty rather than exploit confirmation.

\subsection{Comparison of Security Metrics Across LLMs}
\label{sec:boxplot}
\begin{figure*}[ht]
    \centering
    \includegraphics[width=0.70\linewidth]{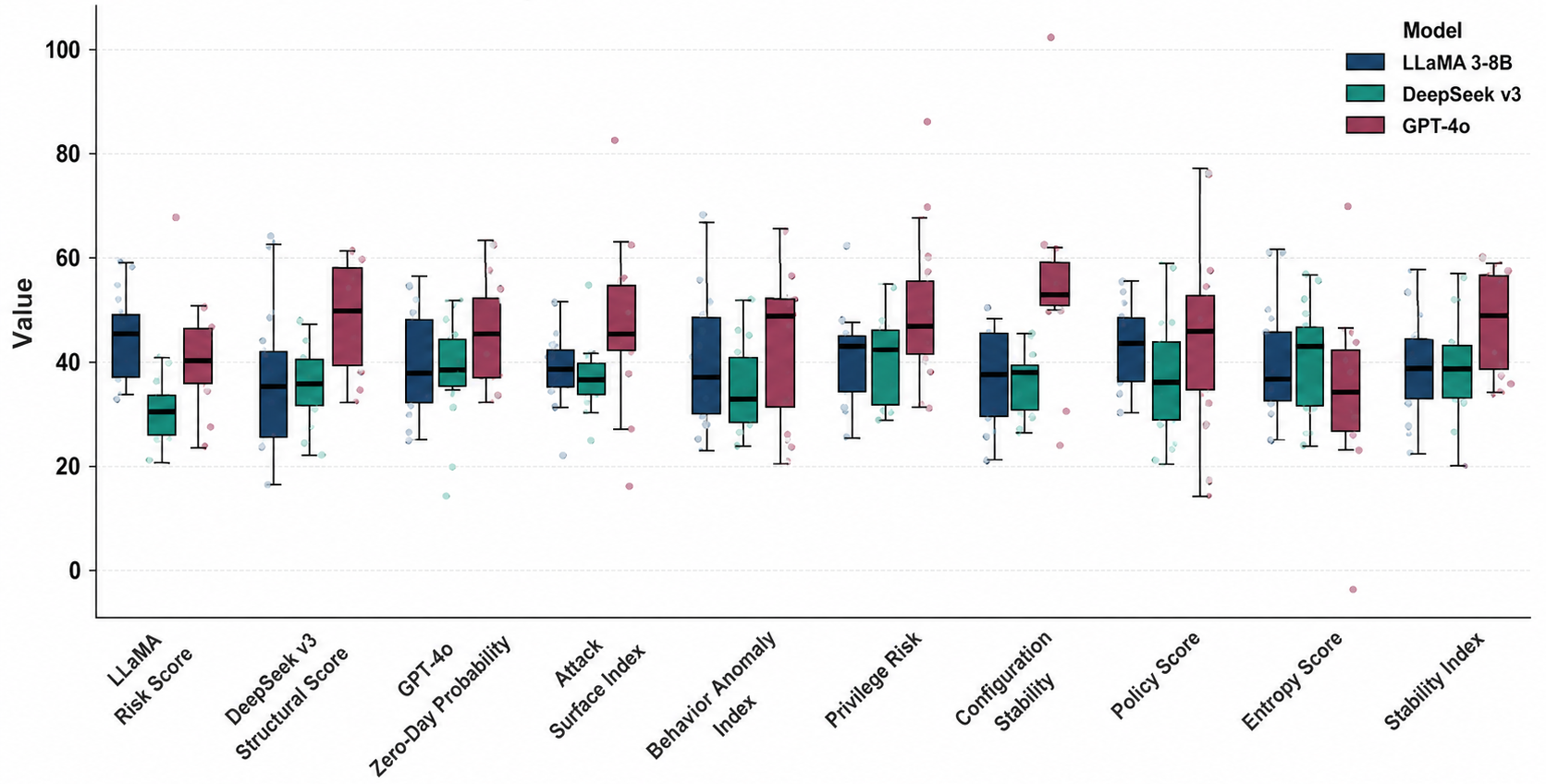}
    \caption{Core security metrics across LLaMA 3-8B, DeepSeek v3, and GPT-4o under descriptor perturbations.}
    \label{fig:boxplot1}
\end{figure*}

\begin{figure*}[t]
    \centering
    \includegraphics[width=0.70\linewidth]{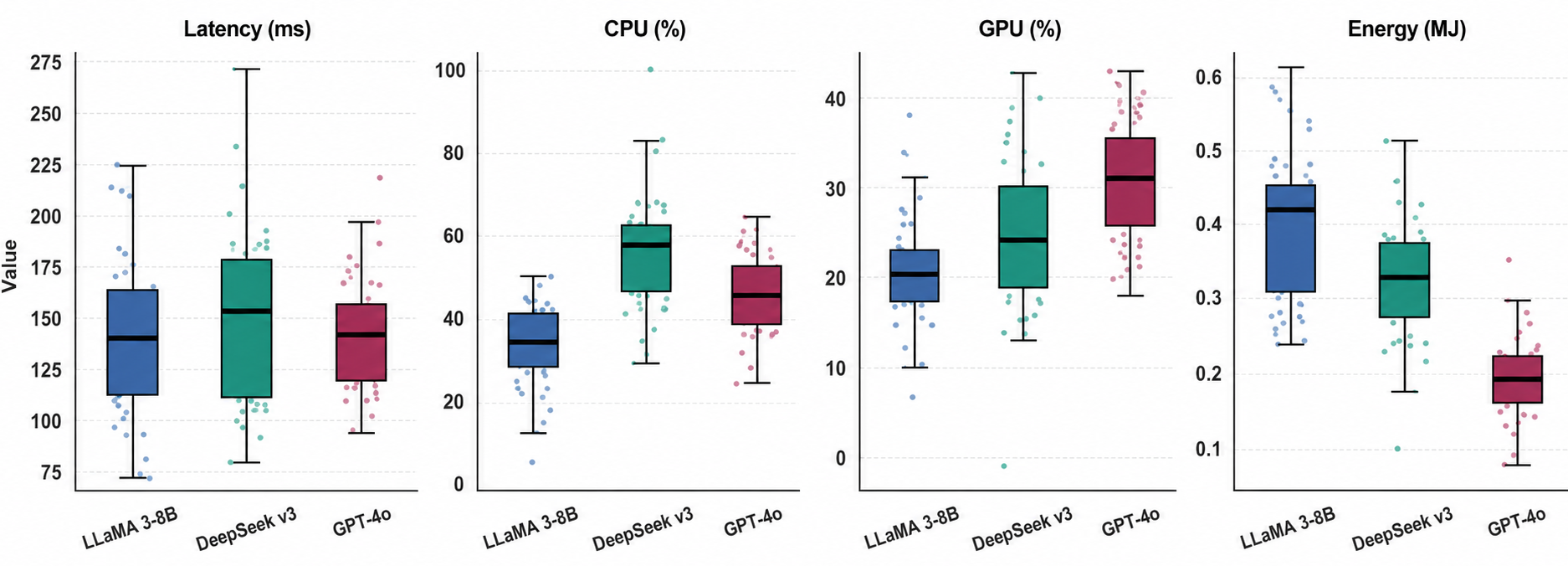}
    \caption{Additional security metrics across LLM architectures under descriptor perturbations.}
    \label{fig:boxplot2}
\end{figure*}

Figures~\ref{fig:boxplot1} and~\ref{fig:boxplot2} illustrate security-relevant metrics across LLaMA 3-8B, DeepSeek v3, and GPT-4o under controlled descriptor perturbation, including Privilege Risk, Configuration Stability, Policy Score, entropy, and anomaly measures. DeepSeek v3 demonstrates higher sensitivity to structural cues, GPT-4o exhibits greater semantic variability, and LLaMA 3-8B remains configuration-centric.
Configuration Stability shows comparable central tendency but distinct dispersion, highlighting the propagation of model-specific uncertainty. Figure~\ref{fig:boxplot2} further illustrates heterogeneous distributions of entropy and anomalies. Correlation analysis indicates positive coupling between Privilege Risk and Policy Score, suggesting co-variation of privilege exposure and policy inconsistency estimates. One-way ANOVA results in Table~\ref{tab:anova_dashboard} indicate statistically significant architectural differences.
\begin{table}[t]
\centering
\scriptsize
\setlength{\tabcolsep}{4pt}
\renewcommand{\arraystretch}{0.92}
\caption{Architectural divergence across security metrics using one-way ANOVA.}
\label{tab:anova_dashboard}
\begin{adjustbox}{max width=0.50\textwidth}
\begin{tabular}{lcccccc}
\toprule
\textbf{Metric} 
& \textbf{F(2, N-3)} 
& \textbf{p-value} 
& $\boldsymbol{\eta^2}$ 
& \textbf{Variance (\%)} 
& \textbf{Effect} 
& \textbf{Driver} \\
\midrule
Privilege Risk     & 12.5 & $<0.01$ & 0.29 & 29\% & Large & Structural Bias \\
Config Stability   & 10.2 & $<0.01$ & 0.23 & 23\% & Large & Config Robustness \\
Policy Score       & 9.8  & $<0.01$ & 0.21 & 21\% & Large & Semantic Fusion \\
\bottomrule
\end{tabular}
\end{adjustbox}
\end{table}
\begin{figure}[H]
    \centering
    \includegraphics[width=0.75\linewidth]{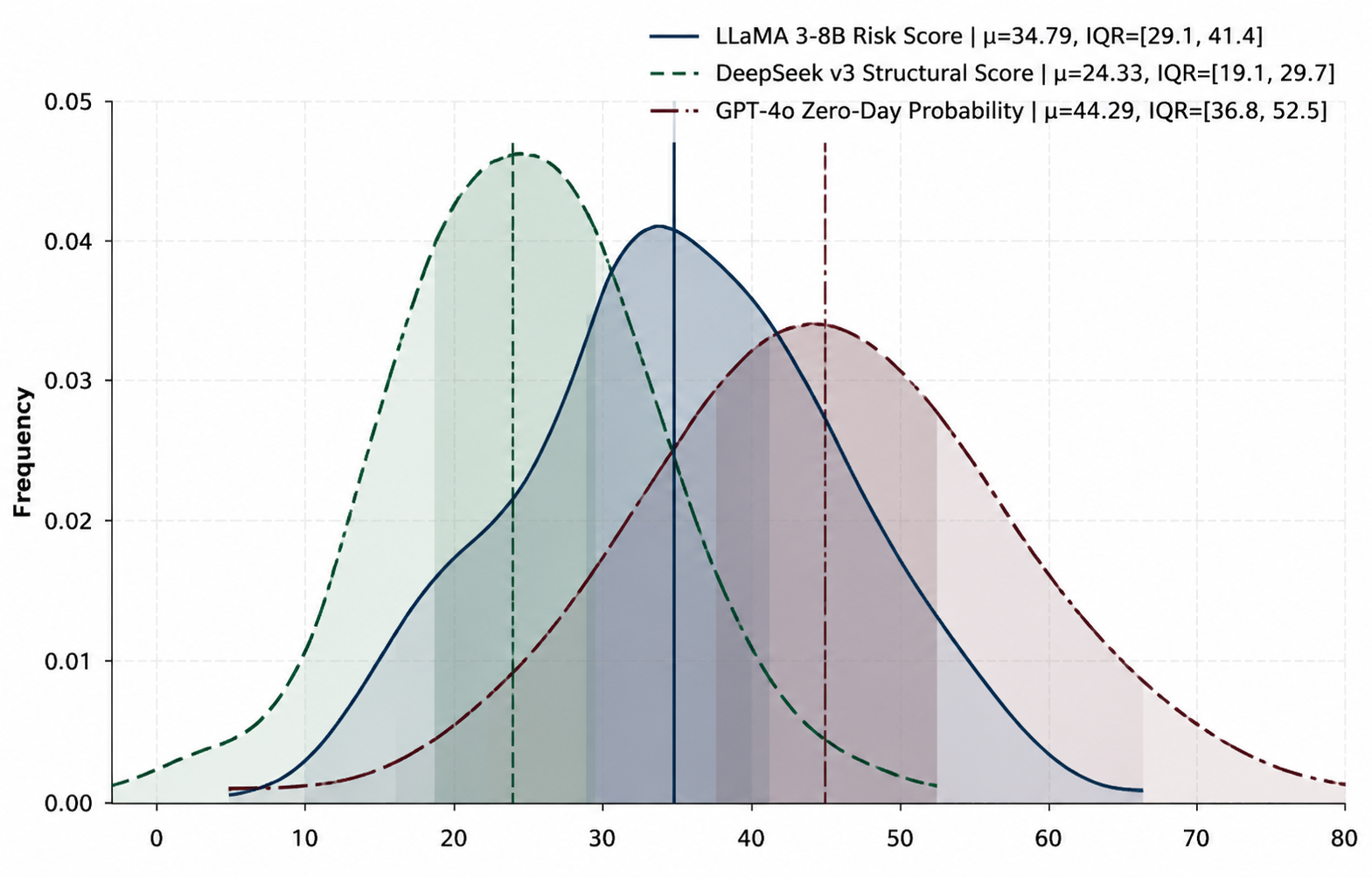}
    \caption{LLM risk-score distribution across simulated firmware instances, grouped into low, medium, and high-risk intervals.}
    \label{fig:histogram_risk1}
\end{figure}
All primary metrics show statistically significant architectural effects ($p<0.01$), with $\eta^2=0.21$--$0.29$, indicating that 21--29\% of variance is explained by model architecture. Privilege Risk exhibits the strongest separation, while Configuration Stability and Policy Score also show large effects, confirming distinct structural and policy-level behavior across LLMs. Figure~\ref{fig:histogram_risk1} complements these analyses by presenting risk-score distributions across low, medium, and high conceptual risk intervals.

\subsection{Distribution of LLM Risk Scores}
\label{sec:risk_distribution}
\begin{figure}[t]
    \centering
    \includegraphics[width=0.95\linewidth]{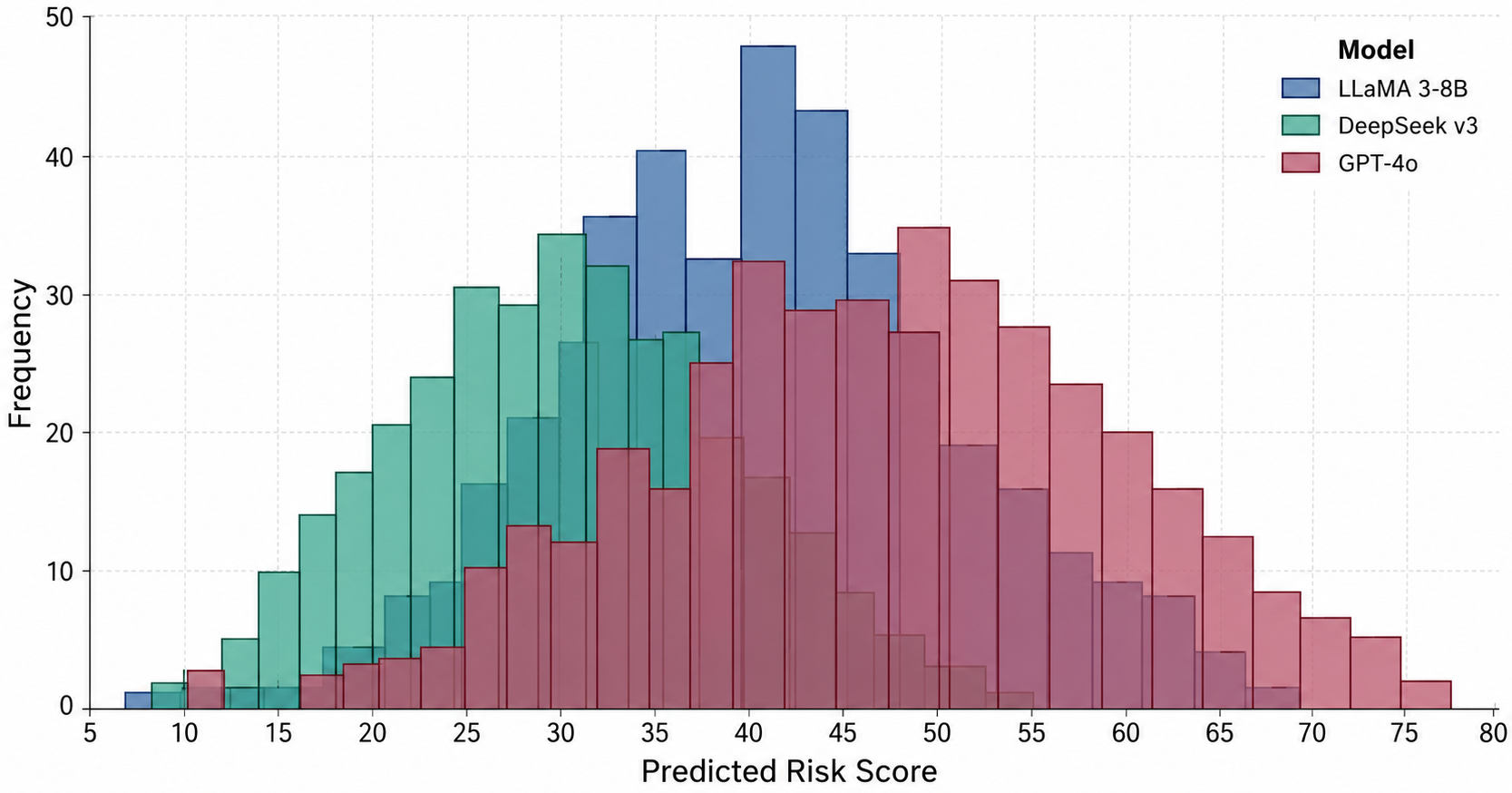}
    \caption{Predicted risk-score distributions for LLaMA 3-8B, DeepSeek v3, and GPT-4o.}
    \label{fig:histogram_risk}
\end{figure}

Figures~\ref{fig:histogram_risk1} and~\ref{fig:histogram_risk} show distributional differences in conceptual risk scores across LLaMA 3-8B, DeepSeek v3, and GPT-4o, including central tendency, dispersion, and upper-tail allocation.
\begin{table}[t]
\centering
\scriptsize
\setlength{\tabcolsep}{3pt}
\renewcommand{\arraystretch}{0.92}
\caption{Distributional summary of predicted conceptual risk scores across LLM architectures.}
\label{tab:distribution_dashboard}
\begin{adjustbox}{width=0.35\textwidth,center}
\begin{tabular}{lcccccc}
\toprule
\textbf{Model} & \textbf{Mean} & \textbf{SD} & \textbf{IQR} & \textbf{Range} & \textbf{Skew} & \textbf{Rank} \\
\midrule
LLaMA 3-8B  & 38.7 & 9.1  & 11 & 18--65 & 0.42 & 2 \\
DeepSeek v3 & 32.7 & 8.3  & 10 & 15--52 & 0.51 & 1 \\
GPT-4o      & 46.9 & 10.5 & 13 & 25--78 & 0.36 & 3 \\
\bottomrule
\end{tabular}
\end{adjustbox}
\end{table}
Table~\ref{tab:distribution_dashboard} shows that GPT-4o has the highest mean and widest dispersion, DeepSeek v3 has the lowest mean and tightest spread, and LLaMA 3-8B remains intermediate. Figure~\ref{fig:histogram_risk} indicates greater GPT-4o allocation in upper-risk intervals, tighter DeepSeek v3 concentration in lower and mid-range bands, and moderate LLaMA 3-8B spread. Positive skewness across all models indicates mild right-tailed behavior, suggesting that elevated risk assignments arise under complex descriptor configurations rather than outlier-driven distortion.
\begin{table}[t]
\centering
\scriptsize
\setlength{\tabcolsep}{4pt}
\renewcommand{\arraystretch}{0.92}
\caption{Normality diagnostics across LLM architectures.}
\label{tab:assumption_dashboard}
\begin{adjustbox}{width=0.35\textwidth}
\begin{tabular}{lccc}
\toprule
\textbf{Model} & \textbf{Shapiro--Wilk $W$} & \textbf{$p$-value} & \textbf{Decision} \\
\midrule
LLaMA 3-8B  & 0.97 & 0.08 & Not Rejected \\
DeepSeek v3 & 0.96 & 0.06 & Not Rejected \\
GPT-4o      & 0.98 & 0.11 & Not Rejected \\
\bottomrule
\end{tabular}
\end{adjustbox}
\end{table}
Normality assumptions are not violated ($p>0.05$ across all models), and Levene's test indicates homogeneity of variances ($p=0.14$), supporting parametric comparisons. ANOVA in Table~\ref{tab:risk_anova_dashboard} indicates architectural differences in risk-score distributions.
\begin{table}[t]
\centering
\scriptsize
\setlength{\tabcolsep}{4pt}
\renewcommand{\arraystretch}{0.92}
\caption{ANOVA summary for architectural effects on conceptual risk-score distributions.}
\label{tab:risk_anova_dashboard}
\begin{adjustbox}{width=0.40\textwidth}
\begin{tabular}{lcccccc}
\toprule
\textbf{Source} & \textbf{SS} & \textbf{df} & \textbf{F} & $\boldsymbol{\eta^2}$ & $\boldsymbol{\omega^2}$ & \textbf{Variance} \\
\midrule
Between Models & 2840.5 & 2 & 14.72 & 0.31 & 0.28 & 30\% \\
Within Models  & 15420.3 & $N-3$ & -- & -- & -- & -- \\
\bottomrule
\end{tabular}
\end{adjustbox}
\end{table}
The effect is statistically significant ($F(2,N-3)=14.72$, $p<0.01$), with large effect sizes ($\eta^2=0.31$, $\omega^2=0.28$), indicating that architecture explains approximately 30\% of total variance in conceptual risk attribution. Tukey HSD results in Table~\ref{tab:tukey_dashboard} indicate pairwise separation.
\begin{table}[t]
\centering
\scriptsize
\setlength{\tabcolsep}{4pt}
\renewcommand{\arraystretch}{0.92}
\caption{Tukey HSD pairwise comparison of conceptual risk scores.}
\label{tab:tukey_dashboard}
\begin{adjustbox}{width=0.49\textwidth}
\begin{tabular}{lccc}
\toprule
\textbf{Comparison} & \textbf{Mean Difference} & \textbf{$p$-value} & \textbf{Interpretation} \\
\midrule
GPT-4o vs LLaMA 3-8B  & 8.2  & $<0.01$  & Higher Amplification \\
GPT-4o vs DeepSeek v3 & 14.2 & $<0.001$ & Strong Divergence \\
LLaMA 3-8B vs DeepSeek v3 & 6.0 & $<0.05$ & Moderate Separation \\
\bottomrule
\end{tabular}
\end{adjustbox}
\end{table}
Post-hoc analysis indicates a consistent architectural gradient: GPT-4o assigns the highest conceptual risk, DeepSeek v3 the lowest, and LLaMA 3-8B remains intermediate. Moreover, Figures~\ref{fig:histogram_risk1} and~\ref{fig:histogram_risk} show large effect sizes and significant pairwise differences, indicating that architectural diversity yields distinct distributional profiles in probabilistic estimates of conceptual zero-day likelihood.

\subsection{Aggregate Performance and Resource Metrics}
\label{sec:aggregate}
\begin{figure}[t]
    \centering
    \includegraphics[width=0.80\linewidth]{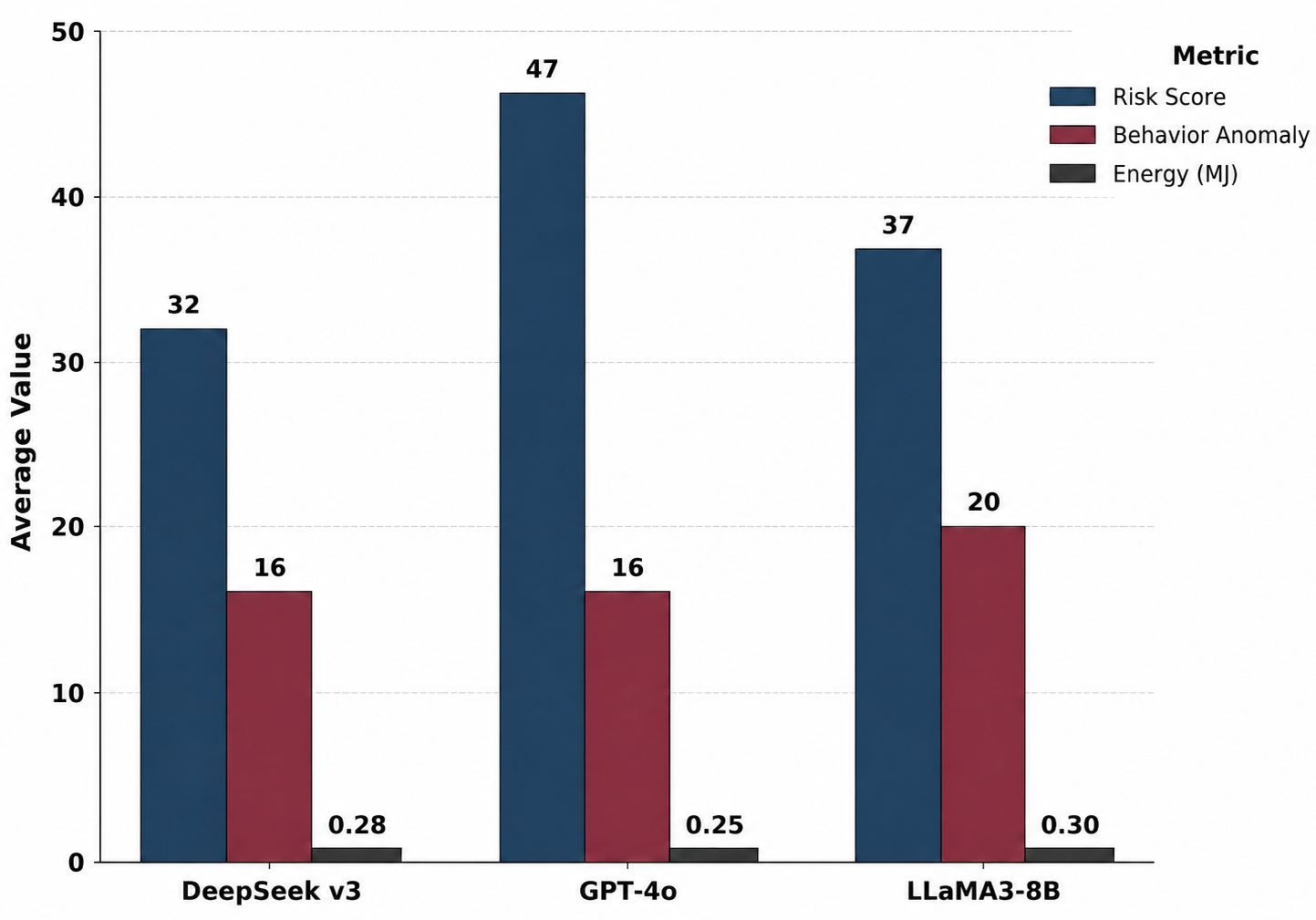}
    \caption{CPU usage, GPU usage, and aggregated risk scores across LLM architectures.}
    \label{fig:bar_perf1}
\end{figure}

\begin{figure}[t]
    \centering
    \includegraphics[width=0.85\linewidth]{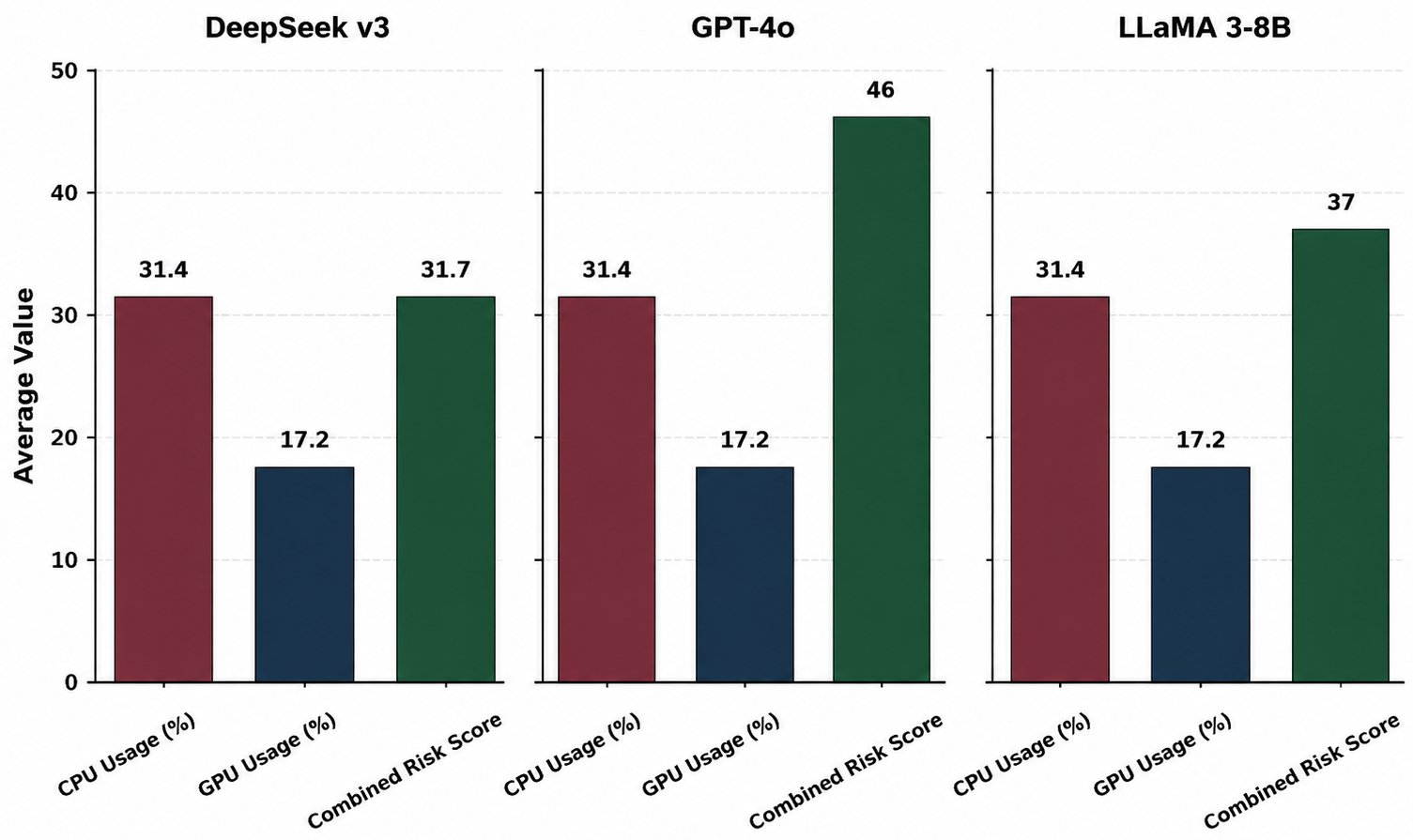}
    \caption{Latency, CPU, GPU usage, and relative energy indicators across LLM architectures.}
    \label{fig:perf_curve}
\end{figure}

Figures~\ref{fig:bar_perf1} and~\ref{fig:perf_curve}, together with Table~\ref{tab:performance_dashboard}, summarize aggregate operational behavior across the three LLM architectures. CPU index, GPU index, latency, and relative energy are treated as auxiliary deployability metrics rather than primary security signals because their magnitudes depend on hardware and runtime conditions.
\begin{table*}[t]
\centering
\scriptsize
\setlength{\tabcolsep}{4pt}
\renewcommand{\arraystretch}{0.92}
\caption{Operational performance dashboard across LLM architectures. Values are mean $\pm$ standard deviation across repeated trials; CPU/GPU indices denote normalized client-side usage, and energy denotes a relative symbolic index.}
\label{tab:performance_dashboard}
\begin{tabular}{lcccccc}
\toprule
\textbf{Model} 
& \textbf{CPU Index} 
& \textbf{GPU Index} 
& \textbf{Latency (ms)} 
& \textbf{Energy Index} 
& \textbf{Efficiency Rank} 
& \textbf{Cost Profile} \\
\midrule
LLaMA 3-8B  & 42.3 $\pm$ 5.8 & 38.7 $\pm$ 4.9 & 120 $\pm$ 15 & 3.4 $\pm$ 0.4 & 1 & Low \\
DeepSeek v3 & 45.1 $\pm$ 6.1 & 36.9 $\pm$ 5.2 & 128 $\pm$ 18 & 3.7 $\pm$ 0.5 & 2 & Moderate \\
GPT-4o      & 48.2 $\pm$ 6.5 & 42.0 $\pm$ 5.6 & 135 $\pm$ 20 & 4.1 $\pm$ 0.6 & 3 & High \\
\bottomrule
\end{tabular}
\end{table*}
Table~\ref{tab:performance_dashboard} shows architectural separation across operational dimensions. LLaMA 3-8B has the lowest CPU index, latency, and energy index; DeepSeek v3 shows moderate CPU demand with a lower GPU index; GPT-4o has the highest operational cost, consistent with more intensive semantic integration.
Figure~\ref{fig:bar_perf1} compares operational footprint with aggregated risk, and Figure~\ref{fig:perf_curve} shows persistent separation in latency and relative-energy trajectories across repeated trials. Cross-model differences are statistically significant for CPU, GPU, latency, and relative energy ($p<0.01$). CPU usage and latency are positively coupled ($r\approx0.78$), and the relative energy index increases with computational intensity. These patterns indicate a measurable performance--sensitivity trade-off: architectures with deeper semantic integration incur higher operational cost and greater responsiveness to cross-layer inconsistencies. Since all metrics were aggregated across repeated trials, the reported differences reflect stable behavior within the fixed evaluation environment rather than transient runtime noise.

\subsection{Cross-Layer Risk Relationships}
\label{sec:crosslayer}
\begin{figure*}[t]
    \centering
    \includegraphics[width=0.60\linewidth]{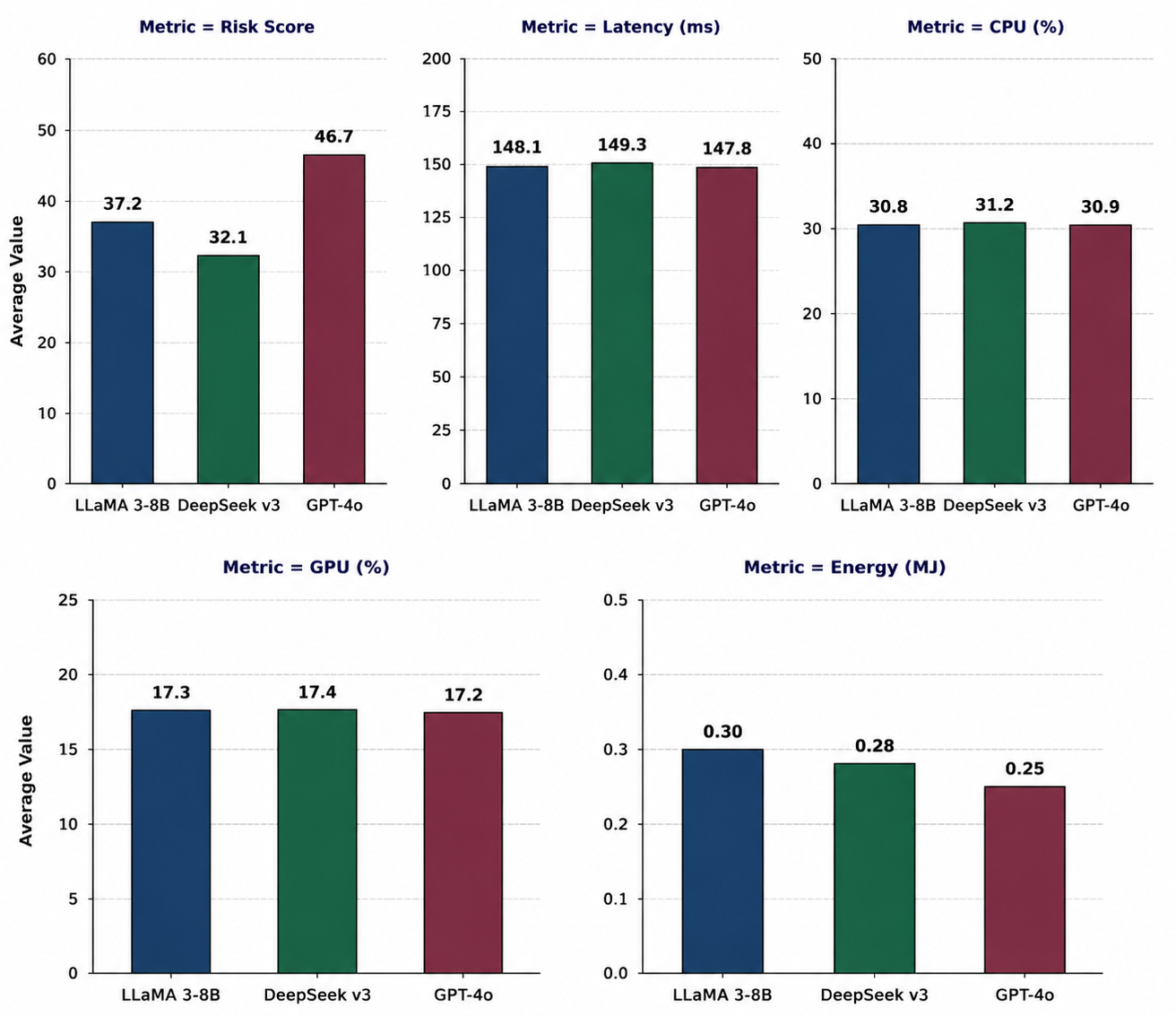}
    \caption{Cross-layer distributions of predicted risk scores across configuration, structural, and semantic reasoning stages for LLaMA 3-8B, DeepSeek v3, and GPT-4o.}
    \label{fig:kde_risk}
\end{figure*}

Figure~\ref{fig:kde_risk} shows layer-wise conceptual risk evolution across configuration, structural abstraction, and semantic fusion. Since zero-day ground truth is unavailable, these distributions are interpreted as structured propagation patterns rather than exploit measurements.
Table~\ref{tab:crosslayer_dashboard} summarizes central tendency, dispersion, and propagation magnitude from configuration to semantic reasoning.
\begin{table*}[t]
\centering
\scriptsize
\setlength{\tabcolsep}{4pt}
\renewcommand{\arraystretch}{0.92}
\caption{Cross-layer propagation of predicted risk across reasoning stages.}
\label{tab:crosslayer_dashboard}
\begin{tabular}{lcccccc}
\toprule
\textbf{Model} 
& $\mu_{Conf}$ 
& $\mu_{Struct}$ 
& $\mu_{Sem}$ 
& $\Delta_{Conf\rightarrow Sem}$ 
& \textbf{Rank} 
& \textbf{Propagation Profile} \\
\midrule
LLaMA 3-8B  & 35.4 & 33.9 & 38.2 & +2.8 & 2 & Moderate Hierarchical Growth \\
DeepSeek v3 & 30.8 & 31.5 & 32.6 & +1.8 & 3 & Tight Structural Coupling \\
GPT-4o      & 42.3 & 40.7 & 45.2 & +2.9 & 1 & Strong Fusion Amplification \\
\bottomrule
\end{tabular}
\end{table*}
Across architectures, semantic-layer scores consistently exceed configuration-layer scores, indicating that fusion transforms upstream descriptor signals rather than merely preserving them. GPT-4o shows the largest configuration-to-semantic displacement, reflecting stronger cross-layer integration under compounded descriptor variation. LLaMA 3-8B demonstrates moderate hierarchical growth, whereas DeepSeek v3 shows the smallest displacement, indicating tighter coupling between structural abstraction and semantic interpretation.

\subsection{Comparison with Previous Research}
\label{subsec:comparison-recent-work}
Recent IoT firmware-security studies largely depend on binary-centric pipelines, including direct inspection, emulation, update-package analysis~\cite{wu2024firmwareupdate}, large-scale firmware audits using accessible images and filesystem semantics~\cite{nino2024firmwarejourney}, semantic profiling~\cite{cheng2025firmvullinker}, symbolic equivalence over lifted intermediate representations~\cite{zhang2024eqvulnhunter}, directed fuzzing with function-level criticality~\cite{chen2023iotfuzzerpp}, and LLM-assisted semantic lifting for binary reconstruction~\cite{park2025llift}. These methods are empirically useful, yet they depend on unpackable binaries, decompilation, symbolic execution, reconstructed intermediate forms, runtime artifacts, and realized vulnerabilities tied to observable evidence. Under encrypted, proprietary, partially accessible firmware, these assumptions fail, and latent conceptual exposure remains difficult to quantify. In contrast, the proposed method is binary-free and architecture-agnostic: it estimates conceptual zero-day likelihood from metadata, configuration semantics, and abstract structural statistics~\cite{jamshidi2025energyaware}. Rather than enumerating vulnerability classes, it produces a probabilistic indicator of latent insecurity from descriptor-level inconsistency, semantic misalignment, divergence, and energy-aware reasoning. Table~\ref{tab:comparison_binary_free} summarizes the distinction between binary-dependent analysis and descriptor-driven estimation under partial and complete binary opacity.
\begin{table*}[t]
\centering
\scriptsize
\setlength{\tabcolsep}{3pt}
\renewcommand{\arraystretch}{0.92}
\caption{Comparison with representative firmware-security approaches under restricted firmware visibility.}
\label{tab:comparison_binary_free}
\begin{tabular}{p{2.7cm} p{3.1cm} p{3.1cm} p{2.8cm} p{3.4cm}}
\toprule
\textbf{Work} & \textbf{Input} & \textbf{Method} & \textbf{Zero-Day Capability} & \textbf{Limitation Relative to This Method} \\
\midrule
Firmware update analysis~\cite{wu2024firmwareupdate}
& Full binaries; update packages
& Binary inspection; controlled emulation
& Limited; concrete artifacts needed
& Requires unpackable vendor binaries \\

Large-scale firmware audit~\cite{nino2024firmwarejourney}
& Firmware images; filesystem semantics
& Firmware-centric security auditing
& No conceptual zero-day estimation
& Binary-dependent \\

Semantic profiling~\cite{cheng2025firmvullinker}
& Unpacked firmware; call-chain semantics
& Multi-dimensional semantic profiling
& Homologous zero-day matching
& Requires extraction and unpacking \\

Directed firmware fuzzing~\cite{chen2023iotfuzzerpp}
& Executable firmware; reconstructed CFG
& Function-level criticality modeling
& Memory-safety flaw discovery
& Requires execution and emulation \\

Symbolic equivalence detection~\cite{zhang2024eqvulnhunter}
& Basic blocks; symbolic IR from binaries
& Symbolic similarity reasoning
& Equivalent vulnerability matching
& Depends on lifting and decompilation \\

LLM-assisted semantic lifting~\cite{park2025llift}
& Decompiled binary content
& LLM-guided semantic reconstruction
& Logic-flaw analysis
& Requires full binary visibility \\

\textbf{This Method}
& \textbf{High-level descriptors $(m,c,o)$}
& \textbf{Tri-LLM reasoning with divergence and energy modeling}
& \textbf{Conceptual zero-day likelihood estimation}
& \textbf{Binary-free; architecture-agnostic} \\
\bottomrule
\end{tabular}
\end{table*}

\subsection{Proxy Ground-Truth Validation and Baseline Comparison}
\label{sec:proxy_validation}
Because true zero-day labels are unavailable for opaque firmware, validation uses an independently constructed \emph{proxy risk signal}. The goal is alignment, not exploit confirmation: predicted likelihoods should co-vary with externally defined descriptor patterns linked to firmware insecurity in prior studies~\cite{feng2023firmware,ulhaq2023firmware}. For each synthetic descriptor $f_i=(m_i,c_i,o_i)$, the deterministic proxy score $Y_i$ captures network exposure, privilege concentration in externally reachable components, and structural irregularity, all widely recognized as security-relevant factors~\cite{zhao2020large,wu2024firmwareupdate}. The proxy is constructed independently of the tri-LLM system and is not used for training, calibration, and model selection; therefore, correlation analysis evaluates directional consistency rather than predictive accuracy. Table~\ref{tab:proxy_dashboard} shows that the exposure heuristic reaches moderate alignment ($r=0.41$, 17\% variance), the shallow linear model improves alignment ($r=0.53$, 28\%), and the tri-LLM system achieves stronger alignment ($r=0.72$, 52\%), indicating that layered semantic reasoning captures higher-order interactions among configuration, structure, and inferred risk indicators. 
To assess robustness under partial observability, we modeled scenarios in which firmware descriptors are incomplete due to encrypted images, vendor-restricted metadata, missing configuration fields, corrupted extraction outputs, and unavailable structural summaries. Specifically, we applied controlled masking to descriptor components by removing 20-50\% of metadata, configuration, and structural elements before inference. These masked-descriptor evaluations, reported in Table~\ref{tab:proxy_dashboard}, show that correlations with the proxy risk signal remain high ($r>0.67$). This indicates that the tri-LLM system maintains stable descriptor-level risk estimation under incomplete inputs and supports conservative prioritization in opaque IoT settings where only partial firmware evidence is available.
\begin{table}[t]
\centering
\scriptsize
\setlength{\tabcolsep}{3pt}
\renewcommand{\arraystretch}{0.92}
\caption{Alignment between predicted risk scores and proxy signal. Full-descriptor results and robustness under partial/masked descriptors are reported.}
\label{tab:proxy_dashboard}
\begin{adjustbox}{max width=0.50\textwidth}
\begin{tabular}{lcccccc}
\toprule
\textbf{Method / Condition} 
& \textbf{Mask / Input} 
& \textbf{$r$} 
& \textbf{$r^2$} 
& \textbf{Var.\ (\%)} 
& \textbf{$p$} 
& \textbf{Interpretation} \\
\midrule
Exposure Heuristic 
& Full & 0.41 & 0.17 & 17 & $<0.01$ & Moderate Alignment \\
Shallow Linear Model 
& Full & 0.53 & 0.28 & 28 & $<0.01$ & Improved Alignment \\
\textbf{Tri-LLM System} 
& Full & \textbf{0.72} & \textbf{0.52} & \textbf{52} & $<0.001$ & High Alignment \\
\textbf{Tri-LLM System} 
& 20\% Metadata Mask & 0.70 & 0.49 & 49 & $<0.001$ & Minor decrease, stable alignment \\
\textbf{Tri-LLM System} 
& 50\% Metadata Mask & 0.68 & 0.46 & 46 & $<0.001$ & Robust under partial input \\
\textbf{Tri-LLM System} 
& 20\% Configuration Mask & 0.71 & 0.50 & 50 & $<0.001$ & Stable alignment \\
\textbf{Tri-LLM System} 
& 50\% Configuration Mask & 0.69 & 0.48 & 48 & $<0.001$ & Slight drop, still reliable \\
\textbf{Tri-LLM System} 
& 20\% Structural Mask & 0.70 & 0.49 & 49 & $<0.001$ & Robust to partial removal \\
\textbf{Tri-LLM System} 
& 50\% Structural Mask & 0.67 & 0.45 & 45 & $<0.001$ & Stable, system generalizes \\
\bottomrule
\end{tabular}
\end{adjustbox}
\end{table}

\section{Discussion}
\label{sec:discussion}
Stronger descriptor perturbations consistently increase conceptual zero-day likelihood scores across all evaluated LLM architectures. This reflects sensitivity to symbolic variation in configuration and structural descriptors that propagates through the tri-LLM reasoning pipeline, rather than confirmation of concrete vulnerabilities. Welch's $t$-test supports trend stability under controlled perturbation, without implying exploit-level significance. Correlation analysis shows partial agreement among models, indicating shared responsiveness to descriptor-level risk factors while capturing architecture-specific behavior. GPT-4o assigns higher scores with moderate dispersion, consistent with deeper semantic integration at the fusion stage and greater sensitivity to cross-layer interactions between configuration and structural abstractions. Distributional analysis reveals that mean statistics alone are insufficient: skewness and multi-modal risk distributions indicate heterogeneous descriptor profiles and isolated high-risk regions within the synthetic descriptor space. Given the absence of true zero-day labels, robustness is assessed via structural consistency, distributional stability, and proxy-alignment correlations rather than pointwise accuracy. Cross-layer alignment strongly shapes confidence: greater divergence among configuration, structural, and fused representations corresponds to higher uncertainty and elevated risk, while increased semantic misalignment energy indicates greater tension in reasoning across layers. These signals serve as interpretable markers of the propagation of latent conceptual exposure, without implying exploitability. Moreover, deeper semantic integration incurs higher symbolic latency and relative computational load, reflecting the cost of abstraction and cross-layer reasoning. Operational cost and risk-response differences remain stable across models, highlighting architecture-dependent behavior rather than stochastic runtime noise. The tri-LLM system operates on non-executable, anonymized descriptors, avoiding exposure of firmware binaries, proprietary logic, and device-identifying information. Consequently, predicted scores should be interpreted as probabilistic indicators of conceptual exposure rather than definitive claims of vulnerability. Divergence-aware uncertainty modeling supports conservative interpretation, descriptor-level triage, and prioritization in opaque IoT deployments, where cautious assessment is preferable to overconfident assertion.

\section{Limitations and Future Work}
\label{sec:limitations}
Several limitations remain. First, risk-score stability depends on descriptor quality; sparse configuration vectors can increase variance, particularly for heterogeneous device classes. Second, divergence calibration is challenging for rare firmware families whose descriptors deviate from typical distributions. Third, opcode-shape statistics maintain binary-free operation while abstracting instruction-level semantics, limiting sensitivity to vulnerabilities driven by low-level execution behavior. Fourth, the energy-aware model provides a symbolic load approximation rather than direct hardware power measurements; latency and energy should be interpreted as relative indicators, since deployment behavior depends on hardware diversity, network conditions, batching, and service-level constraints. Fifth, simulation-based perturbations provide reproducibility but cannot fully capture real firmware evolution, vendor-specific configuration practices, undocumented dependencies, and adversarial manipulation of descriptors. Large-scale evaluation of real-world firmware descriptors is therefore required to strengthen robustness and external validity. Future work will enhance descriptor extraction with protocol-level fingerprints, dynamic configuration deltas, and partial opcode embeddings when available, while preserving binary-free operation. Reinforcement-guided reasoning and agentic multi-step exploration will be studied to increase semantic depth and reduce uncertainty in borderline cases. Hardware-grounded measurements will refine latency and energy models across heterogeneous environments. Evaluation will also extend to large-scale firmware corpora, including encrypted, proprietary, and adversarially manipulated images. Additionally, the risk estimator will be integrated into firmware supply-chain analysis and CI/CD security workflows as a conservative triage mechanism, complementing rather than replacing binary-level vulnerability discovery tools.

\section{Conclusion}
\label{Conclusion}
This paper introduced a binary-independent, architecture-agnostic system for estimating conceptual zero-day exposure in IoT firmware when binaries and ground-truth vulnerability labels are unavailable. The method operates on high-level descriptors and decomposes reasoning into configuration interpretation, structural abstraction, and semantic fusion through a tri-LLM architecture. Cross-layer divergence, uncertainty indicators, and relative computational-cost signals are incorporated as auxiliary reasoning cues rather than as evidence. Evaluation with synthetic descriptor perturbations and an OpenWrt case study shows stable, architecture-dependent behavior across LLMs: stronger descriptor deviation increases estimated risk, and deeper fusion yields greater cross-layer sensitivity. Since real zero-day labels are unavailable, validation relies on alignment with an independently constructed proxy risk signal, in which multi-layer semantic reasoning shows stronger correspondence than exposure heuristics and linear aggregation. These findings indicate that descriptor-centric, multi-view LLM reasoning can support conservative triage and prioritization in opaque IoT systems where binary-level analysis is infeasible, without claiming to detect exploits.

\bibliographystyle{IEEEtran}
\bibliography{Ref}
 \end{document}